\documentstyle[pre,aps,epsf]{revtex}
\draft
\begin{document}
\newcommand{\bm}{\bibitem}
\newcommand{\bgea}{\begin{eqnarray}}
\newcommand{\ndea}{\end{eqnarray}}
\newcommand{\bge}{\begin{equation}}
\newcommand{\nde}{\end{equation}}
\newcommand{\lbl}{\label}
\newcommand{\rf}[1]{(\ref{#1})}
\newcommand{\srf}[1]{\ref{#1}}
\newcommand{\cc}{\mathop{\rm c.c.}\nolimits}
\newcommand{\hf}{\frac{1}{2}}
\newcommand{\bb}[1]{{\bf #1}}
\newcommand{\rr}{\bb r}
\newcommand{\k}{\bb k}
\newcommand{\vfi}{\varphi}
\newcommand{\vna}{\bb \nabla}
\newcommand{\lpl}{{\bb \nabla}^2}
\newcommand{\ibid}{{\it ibid }}
\newcommand{\prtt}{\partial_t}
\newcommand{\prtr}{\partial_r}
\newcommand{\prtf}{\partial_\phi}
\newcommand{\prtpl}{\partial_{\|}}
\newcommand{\prtpp}{\partial_\perp}
\newcommand{\prtR}{\partial_R}
\newcommand{\prtT}{\partial_T}
\newcommand{\prtF}{\partial_\Phi}
\newcommand{\rb}{Rayleigh-B\'{e}nard }
\newcommand{\ob}{Oberbeck-Boussinesq }
\newcommand{\omrm}{\omega_m R_m}
\newcommand{\q}{\bb q}
\newcommand{\hpsi}{\hat{\psi}}
\newcommand{\hrho}{\hat{\rho}}
\newcommand{\hphi}{\hat{\phi}}
\newcommand{\hchi}{\hat{\chi}}
\newcommand{\heta}{\hat{\eta}}
\newcommand{\hj}{\hat{J}}
\newcommand{\bs}{\bar{S}}
\newcommand{\bg}{\bar{g}}
\newcommand{\sprm}{\hskip -5pt \raisebox{2pt}{${}'$}} 
\newcommand{\sprmm}{\hskip -8pt \raisebox{2pt}{${}'$}}
\newcommand{\sprmmm}{\hskip -11pt \raisebox{2pt}{${}'$}}
\newcommand{\pv}[1]{\langle #1\rangle}
\newcommand{\tv}[1]{\overline {#1} } 
\newcommand{\tk}{\tilde{k}}
\newcommand{\gpt}{g_{PT}}
\newcommand{\xis}{\xi_{S}}
\newcommand{\xit}{\xi_2}
\newcommand{\km}{k_{max}}
\newcommand{\calf}{{\cal F}}
\newcommand{\te}{{\epsilon}}
\newcommand{\et}{\epsilon_T}
\newcommand{\stc}{spatiotemporally chaotic }
\title{Phenomenological Theory for Spatiotemporal Chaos in \rb Convection}
\author{Xiao-jun Li$^1$, Hao-wen Xi$^2$ and J. D. Gunton$^1$}
\address{$^1$Department of Physics,
	Lehigh University, Bethlehem, Pennsylvania 18015}
\address{$^2$Department of Physics and Astronomy,
	Bowling Green State University,
	Bowling Green, Ohio 43403}
\date{\today}
\maketitle	
\begin{abstract}
	We present a phenomenological theory for spatiotemporal chaos (STC)
in \rb convection, based on the generalized Swift-Hohenberg
model. We apply a random phase approximation to STC and conjecture a scaling
form for the structure factor $S(k)$ with respect to 
the correlation length $\xit$. We hence obtain 
{\it analytical} results for the time-averaged convective current $J$ and the
time-averaged vorticity current $\Omega$.
We also define power-law behaviors such  as
$J \sim \epsilon^\mu$, $\Omega \sim \epsilon^\lambda$ and 
$\xit \sim \epsilon^{-\nu}$, where $\epsilon$ is the control parameter. 
We find from our theory that $\mu = 1$, $\nu \ge 1/2$ and 
$\lambda = 2 \mu + \nu$ for phase turbulence and that 
$\mu = 1$, $\nu \ge 1/2$ and $\lambda = 2 \mu + 2 \nu$ for 
spiral-defect chaos. These predictions, together with the scaling conjecture 
for $S(k)$, are confirmed by our numerical results. 
Finally we suggest that Porod's law, 
$S(k) \sim 1/\xit k^3$ for large $k$, might be valid in STC. 
\end{abstract}


\section{Introduction} \lbl{int}

	Rayleigh-B\'enard convection (RBC) has long been a
paradigm in the study of 
pattern formation \cite{cr_ho_93,ah_95}. This system consists
of a thin horizontal layer of fluid heated from below. 
There are three dimensionless parameters to describe the system \cite{ch_81}.
The Rayleigh number $R \equiv g \alpha d^3 \Delta T/\kappa \nu$ is the
control parameter,  
in which $g$ is the gravitational acceleration,
$d$ the layer thickness, $\Delta T$ the temperature gradient 
across the layer,  
$\alpha$ the thermal expansion coefficient, $\kappa$ the thermal diffusivity 
and $\nu$ the kinematic viscosity.   
Under the Boussinesq approximation, only the density of the 
fluid is temperature dependent. 
Then the Prandtl number $\sigma \equiv \nu/\kappa$ is all one needs to 
specify the fluid properties.
The third parameter is the aspect ratio 
$\Gamma \equiv L/d$ where $L$ is the horizontal size of the system. 
When the Rayleigh number $R$  surpasses a critical value $R_c$,
the fluid bifurcates from a static conductive state to a convective state,
in which the velocity profile $\bb u = ({\bf u}_\perp, u_z)$ 
and the temperature-deviation profile $\theta$
form certain self-organized patterns.
Those patterns 
also depend on the boundary conditions (b.c.) at the
horizontal surfaces of the container.
The two most studied  b.c. in the literature are the 
{\it rigid-rigid}  b.c., under which  the fluid cannot slip, 
and the {\it free-free}  b.c., 
under which the fluid does not experience any stress. 

The patterns and the corresponding stability domain in RBC have 
been studied extensively
in the classical work of Busse and his coauthors \cite{sc_lo_65,bu_bo_84}
in the $(R, \sigma, k)$ space with $k$ the wavenumber. This stability
domain has hence been  known as the ``Busse balloon'' in the literature.  
For rigid-rigid b.c.,  
the parallel roll state is predicted to be stable inside
the Busse balloon for any given $\sigma$. 
Surprisingly, more 
recent experiments \cite{mo_bo_93,as_st_93} and numerical studies 
\cite{xi_gu_93,be_fa_93,de_pe_94} revealed that for $\sigma \sim O(1)$
and $\Gamma \geq 50$ the parallel roll state yields to a 
spatiotemporally chaotic  state even for states inside 
the Busse balloon. 
This \stc state, called {\it spiral-defect-chaos} (SDC),   
exhibits very complicated dynamics both temporally and spatially
\cite{mo_bo_93,as_st_93}. 
Its discovery has since stimulated many experimental 
\cite{mo_bo_93,as_st_93,mo_bo_96,ca_eg_97}, theoretical \cite{cr_tu_95} 
and numerical \cite{xi_gu_93,be_fa_93,de_pe_94,xi_gu_95} 
efforts to understand it. Despite these efforts, 
few insights have been obtained so far. 

For free-free boundaries,
Zippelius and Siggia  \cite{zi_si_82} 
and Busse and Bolton \cite{bu_bo_84} found that the parallel roll
state is unstable against the skewed-varicose instability immediately above
onset if $\sigma < 0.543$. Busse {\it et al.} \cite{bu_89} further
investigated the dynamics involved and conjectured  
a direct transition from conduction to {\it spatiotemporal chaos}
(STC). This \stc state 
is called {\it phase turbulence} (PT). Recently we reported a large scale
($\Gamma = 60$) numerical simulation of the three dimensional hydrodynamical
equations for $\sigma = 0.5$ under the free-free b.c. \cite{xi_li_97}. From
that simulation, we confirmed the direct transition to PT above onset
and studied various properties of it. The patterns we found 
have very complicated spatial 
and temporal dependences.  

The dynamics of the two types of STC in RBC (SDC and PT) is very complex.
From the theoretical point of view,
it is far from clear whether methods developed in studying
ordered states, such as the Galerkin method \cite{sc_lo_65,bu_bo_84}, the
amplitude equations \cite{se_69}, the phase dynamics \cite{po_ma_79}, etc.,
can be helpful at all. It seems that new concepts and new theoretical tools 
are needed in studying STC \cite{gr_96}. 
We know that 
thermodynamic systems can be characterized to a  certain degree
by {\it global} quantities such as temperature,
pressure, density, etc.. A natural question one may ask is:  
Is it possible to characterize STC by some global quantities? 
The answer to this question is unknown at present but evidence for 
a positive one is very encouraging. 
Recently two of us
found from extensive numerical studies 
that global quantities such as the 
time-averaged vorticity current and the spectrum entropy have 
different behaviors in
the parallel roll and SDC  states \cite{xi_gu_95}.
Furthermore, we demonstrated in our recent paper on PT \cite{xi_li_97}
that both the instantaneous and the time-averaged behaviors 
of global quantities carry valuable information about the state.
Even if global quantities may not be sufficient to 
describe all the dynamical
details of STC, as a first step, we believe that  
knowledge of such quantities will broaden our 
understanding of STC. 

The generalized Swift-Hohenberg (GSH) model of RBC 
\cite{sw_ho_77,cr_80,si_zi_81} 
is widely accepted for theoretical study. 
This model is derived from the three-dimensional hydrodynamic equations,
but is much simpler to study both numerically and analytically. 
After the corresponding $z$-dependences are separated, 
the vertical velocity $u_z(\rr,z,t)$ 
and the temperature-deviation field $\theta(\rr,z,t)$
are reduced to an order parameter $\psi(\rr,t)$ in two-dimensional space 
$\rr = (x,y)$, while 
the vertical vorticity ${\bf e}_z \cdot \vna \times {\bf u}_\perp(\rr, z,t)$ 
is reduced to $\omega_z(\rr,t)$ \cite{sw_ho_77,cr_80,si_zi_81}.  
There are two coupled equations in the GSH model, 
one for $\psi(\rr,t)$ and the other for $\omega_z(\rr,t)$. 
The convective 
patterns in RBC are 
completely determined by the order parameter $\psi(\rr,t)$.
Numerical solutions of this model or its modified versions 
have not only reproduced most patterns observed in 
experiments but also resembled experimental results relatively well
\cite{cr_ho_93,xi_gu_93,be_fa_93,cr_tu_95,xi_gu_95,xi_vi_92}. 
But there are some 
shortcomings in the model \cite{de_pe_94,gr_cr_85}: The stability boundary
of the model does not coincide with that of hydrodynamics; it induces
an unphysical, short-ranged cross roll instability; and both the shape and the
peak position of the power spectrum for SDC are different from those in
the real system. Even so, 
owing to its simplicity and its qualitative resemblance to real systems,
this model is very valuable in studying RBC. 

In this paper we present our {\it analytical} calculations,
using the GSH model,  of the time-averaged
convective current $J = A^{-1} \int d \rr\,\tv{\psi^2(\rr,t)}$ 
and the time-averaged vorticity current 
$\Omega = A^{-1} \int d \rr\,\tv{\omega_z^2(\rr,t)}$ 
in both PT and SDC, 
where $\tv{F(t)}$ represents the time-average of $F(t)$
and $A$ is the area of the system. We carry out our calculations in Fourier
space so the total number of modes $\hpsi(\k,t)$ considered is infinite. 
By assuming the time-averaged two-point correlation function 
$C(\rr_1,\rr_2) \equiv \tv{\psi(\rr_1,t) \psi(\rr_2,t)}/\tv{\psi^2(\rr,t)}$ 
is translation invariant in STC, i.e., $C(\rr_1,\rr_2) = C(\rr_1 - \rr_2)$, 
we find that the phases of two $\hpsi(\k,t)$ fields are 
{\it uncorrelated} in time
unless they have the same wavenumber $\k$. Furthermore, we 
apply a {\it random phase approximation} (RPA) to STC in which
four-point correlation functions are approximated by products of
two-point correlation functions. Using this RPA, we derive 
$J$ and $\Omega$ in  
terms  of the time-averaged and azimuthally averaged structure factor 
$S(k) \equiv \tv{\hpsi^*(\k,t) \hpsi(\k,t)}/J$.
We further assume that $S(k)$ obeys a scaling form
$k S(k) = \xi_2 {\cal F}[(k - \km) \xit]$, in which $\xi_2$ is the two-point
correlation length, ${\cal F}(x)$ is the scaling function and $\km$ is the
peak position of $k S(k)$. Applying this assumption, 
we obtain explicit formulas for both $J$ and $\Omega$. More precisely, we find
that $J = J_0 \epsilon - J_\xi \xit^{-2}$, 
which depends on 
unknown but experimentally measurable parameters, where
$\epsilon = (R - R_c)/R_c$ is the 
reduced control parameter and $J_0$ and $J_\xi$ are both known.
On the other hand, we find that $\Omega = \omega_1 J^2/\xit^2$ for 
rigid-rigid b.c. and $\Omega = \omega_2 w J^2/\xit$ for free-free
b.c., where $w$ is related to the width of the scaling function 
${\cal F}(x)$ and is experimentally measurable. The other coefficients 
$\omega_1$ and $\omega_2$ are known exactly. 
Furthermore, by assuming power law behaviors such that $J \sim \epsilon^\mu$,
$\xit \sim \epsilon^{-\nu}$ and $\Omega \sim \epsilon^\lambda$, we predict
from our theory that $\mu = 1$, $\nu \ge 1/2$ and $\lambda = 2 \mu + 2 \nu$
for rigid-rigid b.c. and
$\mu = 1$, $\nu \ge 1/2$ and $\lambda = 2 \mu + \nu$
for free-free b.c.. This prediction and the scaling assumption for
$S(k)$ have been verified by
our numerical solutions for both PT and SDC. 

Our paper is organized as follows. In Sec. \srf{basic},  
we introduce the GSH model in Fourier space and derive the basic
formulas governing the time-averaged convective current $J$ and 
the time-averaged vorticity current $\Omega$ for any pattern in RBC.
We present a simple theory for PT in Sec. \srf{smpl_th}. This theory
is based on a  conjecture of Busse {\it et al.} \cite{bu_89} 
that PT can be described
by an infinite number of modes lying on a single ring in $k$-space
with different orientations. Although the result for $J$ seems to agree with
our numerical result 
extremely well, this is somewhat accidental since 
it predicts, incorrectly, $\Omega = 0$ and also neglects the strong
couplings between
modes of different $k$'s. In Sec. \srf{rpa} we introduce the RPA of
our theory and use it to calculate explicitly $J$ and $\Omega$
for both PT and SDC. The results are expressed in terms of the structure
factor $S(k)$. 
Sec. \srf{scaling} includes three parts. We first postulate
the scaling form of the structure factor $S(k)$ and expand both 
$J$ and $\Omega$
in the leading order of $1/\xit$. 
We then define power law behaviors for $J$, $\xit$ 
and $\Omega$ in PT and compare the results from 
our theory and our numerical work \cite{xi_li_97}, which agree very well
for both the exponents and the amplitudes. 
In the last part of this 
section
we define power law behaviors for $J$, $\xit$ and $\Omega$ in SDC and
test our theoretical formulas with  our numerical results \cite{xi_li_97_2}. 
The agreement
between our theory and our numerical work
is good in general, except for the amplitude of $\Omega$.
In Sec. \srf{porod} we
discuss the large-$k$ behavior of the structure factor $S(k)$. We conjecture
that Porod's law \cite{po_82,bray}, in which
$S(k) \sim 1/\xis k^3$ for large $k$ with $\xis$
a characteristic length, might be valid for STC.
In the last section, we summarize our results and 
discuss some related issues.

\section{Basic formulas} \lbl{basic}

Near the conduction to convection onset,  the velocity field 
$\bb u(\rr,z,t) = (\bb u_\perp, u_z)$  and
the temperature-deviation field $\theta(\rr,z,t)$ in RBC  
can be approximated by \cite{sw_ho_77,cr_80,si_zi_81}
\bge
	\left[\begin{array}{c}	\bb u_\perp(\rr,z,t)\\ u_z(\rr,z,t) \\ 
		\theta(\rr,z,t) \end{array}\right] \simeq
	\left[\begin{array}{l} 
		u_0(z) \vna \psi(\rr,t) 
		+ \zeta_0(z) \vna \zeta(\rr,t) \times {\bf e}_z \\
		w_0(z) \psi(\rr,t) \\ \theta_0(z) \psi(\rr,t) 
	\end{array} \right],
		\lbl{u_theta_approx}
\nde
where $\vna$ is the gradient operator in two-dimensional space $\rr = (x,y)$. 
For both free-free boundaries at $z=0,1$ 
and rigid-rigid boundaries at $z = \pm \hf$, 
the explicit forms of $u_0(z)$, $w_0(z)$ and $\theta_0(z)$ are  
given in Ref. \cite{cr_80}; 
one takes $\zeta_0(z) = 1$ for free-free boundaries and 
$\zeta_0(z) = z^2 -\frac{1}{4}$ for rigid-rigid boundaries.
Notice that the vertical vorticity 
${\bf e}_z \cdot \vna \times \bb u_\perp$ is now replaced by
$\zeta_0(z) \omega_z(\rr,t)$ in which $\omega_z(\rr,t) = - \lpl \zeta(\rr,t)$. 
Inserting Eq. \rf{u_theta_approx} into 
the three-dimensional hydrodynamical equations in RBC and applying a few more
approximations,  one ends up with the two-dimensional generalized 
Swift-Hohenberg (GSH) model of RBC  \cite{sw_ho_77,cr_80,si_zi_81}. 
Although some of the approximations are not systematic,
the amplitude equations for the GSH model and the hydrodynamical equations
are the same in the leading order near onset. 
There are two coupled equations in the GSH model,
one for the order parameter  $\psi(\rr,t)$ and the other for the
mean-flow field $\zeta(\rr,t)$. 
The convective patterns are completely determined by the
order parameter $\psi(\rr,t)$. The GSH model has been proven very successful
in characterizing convective patterns under quite broad conditions 
\cite{cr_ho_93}. 

In the GSH model,
the order parameter $\psi(\rr,t)$ satisfies \cite{sw_ho_77,cr_80,si_zi_81}  
\bge
        \tau_0\left[\prtt \psi + \bb U \cdot \bb \nabla \psi\right] 
        = \left[\epsilon -(\xi_0^2/4 k_c^2)({\bb \nabla}^2 
		+ k_c^2)^2 \right] \psi
               - N[\psi].
                                \lbl{gsh}
\nde
Here $N[\psi]$ is the nonlinear term to be specified soon and 
$\bb U(\rr)$ is the mean flow velocity given by
$\bb U(\rr) = \vna \zeta(\rr,t) \times {\bb e}_z$,  in which  
\bge
        \left[\prtt -\sigma (\lpl - c^2)\right] \lpl \zeta
                = g_m {\bb e}_z \cdot \left[ \vna(\lpl \psi)
                \times \vna \psi\right]. \lbl{mf}
\nde
In the GSH equations, the reduced Rayleigh number $\epsilon \equiv (R/R_c) -1$
is the control parameter, in which  $R$ and $R_c$ are
the Rayleigh number and its critical value at onset.
The Prandtl number $\sigma$ parameterizes the fluid.
While $k_c$ is the critical wavenumber at onset, the other parameters
model the properties of the system. The values of these parameters depend
on the boundary conditions (b.c.), more precisely \cite{cr_80,para},
\bgea
	{\rm for \ free-free \ b.c.: \quad} 
	&& R_c = 27 \pi^4/4, \  k_c = \pi/\sqrt{2},
	\  \tau_0 = 2 (1 + \sigma^{-1})/3 \pi^2, \nonumber \\
	&& \xi_0^2 = 8/3 \pi^2,
	\   g_m = 6, \ c^2 = 0; \lbl{para_fr_fr} \\
	{\rm for \ rigid-rigid \ b.c.: \quad} 
	&& R_c = 1707.762, \  k_c = 3.117, 
	\  \tau_0 = (1 + 0.5117 \sigma^{-1})/19.65, \nonumber \\
	&& \xi_0^2 = 0.148, 
	\   g_m = 24.77, \ c^2 = 10. \lbl{para_rg_rg} 
\ndea
	
	It is easier to analyze the GSH equations theoretically
in Fourier space than in real space.
By convention, we define the Fourier transformation and 
its inverse transformation of an 
arbitrary function $F(\rr)$ as 
\bge
	\hat{F}(\k) = \frac{1}{A} \int d \rr \, e^{- i \k \cdot \rr} F(\rr)
	\quad {\rm and} \quad 
	F(\rr) = \sum_{\bb k} \hat{F}(\k) e^{i \k \cdot \rr},
		\lbl{ft}
\nde
where $A$ is the area of the system. 
Note that $\hat{F}^*(\k) = \hat{F}(-\k)$ for any real function $F(\rr)$.
It is easy to check that Eq. \rf{gsh} can be rewritten in Fourier space as 
\bge
	\tau_0\prtt \hat{\psi}(\k) 
	 + \hat{V}(\k)
	= r(\epsilon; k) 
 	 \hat{\psi}(\k) -\hat{N}(\k), 
	\lbl{ftgsh}
\nde
where $V(\rr) = \tau_0 \bb U \cdot \vna \psi$ and 
\bge
	r(\epsilon; k) = \epsilon - \xi_0^2 (k^2 - k_c^2)^2/4 k^2_c.
		\lbl{r}
\nde
Since $\psi(\rr,t)$ is real, one has $\hpsi^*(\k,t) = \hpsi(-\k,t)$. 
The nonlinear $\hat{N}(\k)$ term has been  evaluated at onset \cite{cr_80},
\bge
	\hat{N}(\k) = \sum_{\k_2, \k_3} 
			g(\hat{\k} \cdot \hat{\k}_2)
			\hat{\psi}^*(\k_2) \hpsi(\k_3) \hpsi(\k+\k_2-\k_3),
		\lbl{nl}
\nde
where the coupling constant $g(\cos \alpha)$ is 
given in Ref. \cite{cr_80} with $\alpha$ the angle between $\k$ and 
$\k_2$. Rigorously speaking, the exact forms of Eqs. \rf{gsh}, \rf{mf} and
\rf{nl} are derived near onset and deviations from them in real physical
systems are possible for large enough $\epsilon$. 
But we disregard such complexity and take them 
as our model for further study.  

One may take an adiabatic approximation $(\partial_t = 0)$ in Eq. \rf{mf}
by neglecting the first term on the left-hand side.
This term is small in comparison with the other terms, which 
can be verified by applying the same perturbation as that in 
phase dynamics \cite{po_ma_79}. With this approximation, 
it now is easy to solve Eq. \rf{mf} for $\hat{\zeta}(\k)$,
which indicates that the mean-flow field is slaved by the $\psi(\rr,t)$ 
field. We are also interested in the 
vertical vorticity $\omega_z(\rr) = - \lpl \zeta(\rr)$.   
From Eq. \rf{mf}, it is straightforward to get that 
\bge
	\hat{\omega}_z(\k) = k^2 \hat{\zeta}(\k)
		= \sum_{\k_2} f(\k; \k_2) \hpsi(\k_2) \hpsi(\k - \k_2),
	\lbl{vtct}
\nde
where, with an exchange of index $\k_2 \to \k - \k_2$, 
\bge
	f(\k; \k_2) = \frac{g_m}{2 \sigma (k^2 + c^2)} (k^2 - 2 \k \cdot \k_2) 
		({\bb e}_z \cdot \k_2 \times \k). \lbl{cplf}
\nde
Applying these results, one may easily 
evaluate the mean-flow contribution to Eq. \rf{ftgsh}, which is given by
\bge
	\hat{V}(\k) = \sum_{\k_2, \k_3}   v(\k; \k_2; \k_3) 
			\hat{\psi}^*(\k_2) \hpsi(\k_3) \hpsi(\k + \k_2 -\k_3),
			\lbl{ftmf}
\nde
where
\bge
	v(\k; \k_2; \k_3) = \frac{g_m \tau_0}{2 \sigma} 
		\frac{[{\bb e}_z \cdot \k \times (\k_3 - \k_2)]
			[{\bb e}_z \cdot \k_3 \times \k_2] (k_2^2 -k_3^2)}
		{|\k_3 -\k_2|^4 + c^2 |\k_3 - \k_2|^2} \, .
			\lbl{cplv}
\nde
Notice that the coupling constant $v(\k;\k_2;\k_3)$ is zero under two
conditions: (1) If all $\k$ allowed in $\hpsi(\k)$ point at one single 
direction, say $\hat{\k}$; 
or, (2) if all $\k$ lie on one single ring, say $|\k| = k$. For this reason, 
ordered states  such as parallel rolls, hexagons, 
concentric rings, etc., do not have significant mean-flow couplings.  
Furthermore, the coupling constant $v(\k;\k_2;\k_3)$ 
seems to have a pole at $\k_2 = \k_3$.
The real situation, however, is more subtle. Assume that 
$\k_3 = \k_2 + \q$ with $\q$ very small; then
$v \sim ({\bb e}_z \cdot \k \times \q) 
({\bb e}_z \cdot \k_2 \times \q)(\k_2 \cdot \q)
/(c^2 q^2 + q^4)$. For rigid-rigid boundaries ($c^2 = 10$),
there is no pole (since $v \sim q$) at $q = 0$. But for
free-free boundaries ($c^2 = 0$), 
a pole normally exists (since $v \sim 1/q$)
unless $\q \| \k$, or $\q \| \k_2$, or $\q \perp \k_2$.

	In this paper, we will mainly focus on two global quantities: One 
is the total convective current defined by
\bge
	J(t) = \frac{1}{A} \int d \rr \, \psi^2(\rr, t)
		= \sum_{\k} \hat{J}(\k,t) \quad {\rm with} \quad
		\hat{J}(\k,t) = \hpsi^*(\k,t) \hpsi(\k,t); 
		\lbl{cvcrnt}
\nde
the other is the total vorticity ``current'' defined by
\bgea
	\Omega(t) &=& \frac{1}{A} \int d \rr \, \omega_z^2(\rr, t) \nonumber \\
		&=& \sum_{\k_1,\k_2,\k_3,\k_4}  f(\k_1+\k_2;\k_2) 
		f(\k_1+\k_2;\k_3)
		 \hpsi^*(\k_1) \hpsi^*(\k_2) \hpsi(\k_3) \hpsi(\k_4) \delta_{k_1+\k_2,\k_3+\k_4}.
		\lbl{vtctcrnt}
\ndea
Notice that 
$f(\k_1+\k_2;\k_2) \sim (k_1^2 -k_2^2)({\bb e}_z \cdot \k_2 \times \k_1)$. 
So $\Omega(t) =0$
if all the wavenumbers allowed in $\hpsi(\k)$ 
point at one single direction $\pm \hat{\k}$  
or lie on one single ring $|\k| = k$. 
In other words, the vorticity current must be generated by couplings
between modes of different $k$ and $\hat{\k}$.
From Eq. \rf{ftgsh}, it is easy to derive that
\bgea
	\tau_0 \prtt \hj(\k,t) &=& 2 r(\epsilon; k) \hj(\k,t)\nonumber \\
		&-& \sum_{\k_2,\k_3}
		\left[g(\hat{\k} \cdot \hat{\k}_2)+v(\k; \k_2; \k_3)\right]
		\left[\hpsi^*(\k) \hpsi^*(\k_2) \hpsi(\k_3) \hpsi(\k+\k_2-\k_3)
		+ \cc \right].
		\lbl{keqcvcrnt}
\ndea
In principle, this is the equation determining the structure of the
convective current $\hj(\k,t)$ which, however, 
is beyond our present goal. 
Now applying the relations $\hpsi^*(\k) = \hpsi(-\k)$ and 
$v(\k; \k_2; \k_3)=-v(\k; \k_3;\k_2)$, and, 
exchanging the summation indices $\k \to -\k$, $\k_{2,3} \to - \k_{2,3}$ 
for the
$g$ terms and  $\k \to \k + \k_2 -\k_3$, $\k_2 \leftrightarrow \k_3$ for  
the $v$ terms, one obtains from the above equation and the definition of
$J(t)$ that 
\bge
	\hf \tau_0 \prtt J(t) = \sum_{\k} r(\epsilon; k) \hj(\k,t)
		- \sum_{\k_1, \k_2,\k_3} g(\hat{\k}_1 \cdot \hat{\k}_2)
		\hpsi^*(\k_1) \hpsi^*(\k_2) \hpsi(\k_3) \hpsi(\k_1+\k_2-\k_3).
		\lbl{eqcvcrnt}
\nde
This is the equation determining the total convective 
current $J(t)$. 
Notice that the $v$ terms vanish from this equation, which 
can also be derived directly from Eq. \rf{gsh}  
by converting the corresponding integral in Eq. \rf{cvcrnt} into 
a surface term. In general, the $v$ couplings affect 
$J(t)$ implicitly by modifying  its structure $\hj(\k,t)$ 
unless, of course,  $v \equiv 0$.  

	For stationary states, the convective current and the vorticity
current are time-independent. This, however, is no long true if the state
is spatiotemporal chaotic.  For a spatiotemporal chaotic state, these
two currents  normally fluctuate in time around some well-defined averaged 
values: see Refs. \cite{xi_gu_95,xi_li_97}.
While the fluctuations appear chaotic in time, they are 
relatively small in comparison with their averaged values. For simplicity, we
only consider the two corresponding time-averaged currents in our theory. 
We now introduce the time-average operator $\cal T$ defined by 
\bge
	{\cal T} F(t) \equiv \tv {F(t)} 
	= \lim_{T \to +\infty} \frac{1}{T} \int^T_0 dt F(t).
		\lbl{tv}
\nde
Applying $\cal T$ to Eq. \rf{eqcvcrnt} yields
\bge
	\sum_{\k} r(\epsilon; k) \tv{\hj(\k,t)}
		- \sum_{\k_1, \k_2,\k_3,\k_4} g(\hat{\k}_1 \cdot \hat{\k}_2)
		\tv{\hpsi^*(\k_1) \hpsi^*(\k_2) \hpsi(\k_3) \hpsi(\k_4)} \,\delta_{\k_1+\k_2,\k_3+\k_4} = 0.
		\lbl{tveqcvcrnt}
\nde
In the next several sections, we show how, under various assumptions,  
to calculate the time-averaged convective current of STC 
from this equation. The time-averaged vorticity current can be obtained
with $\cal T$ acting on Eq. \rf{vtctcrnt}. For simplicity, 
we denote from now on
$\hj(\k) = \tv{\hj(\k,t)}$, $J = \tv{J(t)}$ and $\Omega = \tv{\Omega(t)}$.

	Finally we introduce the time-averaged structure factor defined by  
\bge
	\hat{S}(\k) = \hj(\k)/J \quad
		{\rm with} \quad 
	\sum_{\k} \hat{S}(\k) = 1,
		\lbl{dfs}
\nde
and the corresponding averages
\bge
	\pv{\hat{F}}_\k 
		= \sum_{\k} \hat{S}(\k) \hat{F}(\k). \lbl{sv}
\nde
With this notation, the first term in Eq. \rf{tveqcvcrnt} can be rewritten as
$\pv {r(\epsilon)}_\k J$. If the $k$-dependence and the angular dependence 
in $\hat{S}(\k)$ can be separated, then it is more convenient to define
\bge
	\hat{S}(\k) = (2 \pi)^2 A^{-1} S(k) \Phi(\alpha) \quad {\rm with} 
	\quad \int^\infty_0 d k \, k S(k) = 1 \quad {\rm and} \quad 
	\int^{2 \pi}_0 d \alpha \, \Phi(\alpha) = 1, \lbl{dfssep}
\nde
where $\alpha$ is the angle between $\k$ and some reference direction. 
Here 
the discrete $\k$ lattice has been converted 
into a continuous one. So a proper phase factor has been taken into account. 
Notice also that $\Phi(\pi + \alpha) = \Phi(\alpha)$ since 
$\hat{S}(-\k) = \hat{S}(\k)$. 
Now the corresponding averages with respect to $S(k)$ and $\Phi(\alpha)$
are defined as 
\bge
	\pv F_k = \int^\infty_0 d k \, k S(k) F(k) \quad {\rm and}
	\quad \pv F_\alpha = \int^{2 \pi}_0 d \alpha \, \Phi(\alpha) F(\alpha).
	\lbl{ave}
\nde
For $\hat{F}(\k) = F(k,\alpha)$, it is easy to see that 
$\pv{\hat{F}}_\k = \pv{F}_{k,\alpha}$ if a separation like 
Eq. \rf{dfssep} holds.

\section{A Simple Theory for PT} \lbl{smpl_th}

	In order to calculate the time-averaged convective current and
the time-averaged vorticity current from 
Eqs. \rf{tveqcvcrnt} and \rf{vtctcrnt},
it is obvious that more information about the corresponding state is needed.
We now present a simple model for PT, following a conjecture by Busse
and his coauthors \cite{bu_89}, and calculate 
the corresponding convective current. For this task, we notice 
first that a PT state 
consists of many superimposed rolls with different orientations
\cite{bu_89,xi_li_97}, among which  no particular direction is preferred.
On the other hand, 
since this PT occurs immediately above the onset 
\cite{bu_bo_84,zi_si_82,bu_89,xi_li_97},
the amplitude of the wave number $k$ lies in the vicinity of $k_c$. 
Our numerical simulations \cite{xi_li_97} also revealed that the time-averaged
structure factor is isotropic azimuthally.

As a first
attempt, we assume that a PT state is composed of many parallel rolls, whose
wave numbers lie on a ring and whose amplitudes are equal. 
More precisely, we assume that 
\bge
	\hpsi(\k,t) = \psi_0(t)\,\delta_{k,q} \sum^M_{i = 1} 
	e^{i \phi(\beta_i,t)} \delta_{\alpha, \beta_i(t)}\,,  \lbl{ring}
\nde
where  $\psi_0(t)$ and $\phi(\beta_i,t)$ 
are the amplitude and the phases 
of those selected modes, 
$q (\simeq k_c)$ and $\beta_i(t)$ are the amplitude and the angles 
of their corresponding wave numbers, while $M$ is the total number of those 
modes, and 
$\alpha$ is the angle
between $\k$ and some reference direction.
From Eq. \rf{cvcrnt}, one finds that 
$J(t) = \sum_{\k} |\hpsi(\k,t)|^2 = M \psi^2_0(t)$.   
One has $M=2$, $\psi_0 = \sqrt{J/2}$, $\phi(\beta_i,t) = {\rm const.}$,  
and $\beta_i = i \pi$ with $i = 0,1$ for parallel rolls, and 
$M=6$, $\psi_0 = \sqrt{J/6}$, $\phi(\beta_i,t) = {\rm const.}$,  
and $\beta_i = i \pi/3$ with $i = 0, 1, \cdots, 5$ 
for hexagons \cite{cr_80}.  
For PT, we take $M \to + \infty$.  
Since $J(t)$ in PT has a well-defined time-averaged value
with small fluctuations \cite{xi_li_97}, we expect the same behavior for 
$\psi_0(t)$. But we speculate that  both the phases $\phi(\beta_i,t)$
and  the angular distribution $\{\beta_i(t)\}$
are irregular in time, which leads to 
the spatiotemporal chaotic behavior in PT.
From $\hpsi^*(\k,t) = \hpsi(-\k,t)$, one finds that 
if $\beta_i(t)$ is selected, so must be $\pi+\beta_i(t)$ with
$\phi(\pi + \beta_i, t) = - \phi(\beta_i,t)$.

	We now use this model to calculate the time-averaged convective 
current from Eq. \rf{tveqcvcrnt}. While the first term can be easily evaluated as
$r(\epsilon,q) J$, the second term is much more complicated. Notice that the
condition $\k_1 + \k_2 = \k_3 + \k_4$ imposes a very strong constraint on the
available wave numbers on a ring. 
There are only three possibilities for the condition to be satisfied, see Fig. \ref{fig_ring}:  
(a) If $\k_1 + \k_2 = 0$, then
$\k_3 + \k_4 = 0$; (b) if $\k_1 \neq \k_2$ and $\k_1 + \k_2 \neq 0$, then
either $\k_3 = \k_1$ and $\k_4 = \k_2$ or $\k_3 = \k_2$ and $\k_4 = \k_1$;
or, (c) if $\k_1 = \k_2$, then $\k_3 = \k_4 = \k_1$. 
It is more convenient to express these constraints in terms of their angles,
which can be summarized, correspondingly, as: 
(a) $\delta_{\alpha_2,\alpha_1+\pi} \delta_{\alpha_4,\alpha_3+\pi}$,
(b) $(1 - \delta_{\alpha_2,\alpha_1+\pi} - \delta_{\alpha_2,\alpha_1})
	(\delta_{\alpha_3,\alpha_1} \delta_{\alpha_4,\alpha_2}
	+\delta_{\alpha_3,\alpha_2} \delta_{\alpha_4,\alpha_1})$,
and (c) $\delta_{\alpha_2,\alpha_1} \delta_{\alpha_3,\alpha_1}
	\delta_{\alpha_4,\alpha_1}$.
Now inserting Eq. \rf{ring} into the second term of Eq. \rf{tveqcvcrnt}
and applying $\phi(\beta_i + \pi) = - \phi(\beta_i,t)$ and these constraints, 
after some algebra, one finds that the second term is simply
$- M^2 g_M \tv{\psi^4_0(t)}$ 
with
\bge
	g_M = g(-1) (1 - \frac{2}{M}) - \frac{1}{M} g(1) + \frac{2}{M^2} 
	\sum^M_{i,j=1} \tv{g[\cos(\beta_i(t) - \beta_j(t))]}. \lbl{gm}
\nde
Here $\tv{\psi_0^4(t)}$ and $\tv{g[\cos(\beta_i(t) - \beta_j(t))]}$ have been
decoupled, which seems reasonable.
Since $J = M \tv{\psi^2_0(t)}$, if one neglects the fluctuations of 
$\psi_0(t)$, one has then $M^2 \tv{\psi^4(t)} \simeq J^2$. 
From Eq. \rf{tveqcvcrnt}, this leads to the solution for the total 
time-averaged convective current
\bge
	J \simeq r(\epsilon,q)/g_M, \lbl{tvcvcrnt_pt}
\nde
in addition to the conduction solution $J = 0$.
This solution 
reproduces the known results \cite{cr_80} for both parallel rolls
with $g_2 = g(-1) + \hf g(1)$ and hexagons with
$g_6 = \frac{1}{6}[6 g(-1) + 4 g(-\hf) + 4 g(\hf) + g(1)]$. 

	For PT, we expect that $\beta_i(t)$ and $\beta_j(t)$ are uncorrelated
in time and that fluctuations of $\beta_i(t)$ can be neglected such that
$\tv{\beta_i^n(t)} \simeq \tv{\beta_i(t)}^n$ for any positive integer $n$. 
So we may 
make another approximation such that
$\tv{g[\cos(\beta_i(t) - \beta_j(t))]} 
\simeq g[\cos(\tv{\beta_i(t)} - \tv{\beta_j(t)})]$, where $\{\tv{\beta_i(t)}\}$
should be uniformly distributed between $[0, 2 \pi]$. From this approximation
and $\sum_{i = 1}^M \to (M/2 \pi) \int^{2 \pi}_0 d \alpha$ as $M \to +\infty$,
one finds that
\bge
	g_\infty = g(-1) + \frac{2}{\pi} \int^\pi_0 d \alpha\,g(\cos\alpha).
	\lbl{g_inf}
\nde
Using the explicit formula given in Ref. \cite{cr_80} for free-free
boundaries, one has finally that
\bge
	g_{PT} = 0.855922 + 0.0458144 \sigma^{-1} + 0.0709326 \sigma^{-2},
	\lbl{g_pt}
\nde
where $\sigma$ is the Prandtl number. Since typically 
$q \simeq k_c(1 +q_0 \epsilon)$, one finds that the time-averaged convective
current $J \simeq r(\epsilon,q)/g_{\rm PT} \simeq \epsilon/\gpt$ in PT, 
recalling
Eq. \rf{r}. For $\sigma = 0.5$, this simple theory gives 
$\gpt \simeq 1.2313$. In comparison,
we found $\gpt \simeq 1.27 \pm 0.03$ from our three-dimensional numerical 
calculations \cite{xi_li_97}. Considering all the approximations we have made,
such a good agreement is very encouraging. 

However, this
simple model apparently misses two important features of PT. The first
is the lack of the vorticity current. Since all $\k$ lie on a single ring
in our model, the vorticity current is identically zero from our discussions 
in the previous section. This, however, 
is not born out by our numerical calculations. Secondly, 
the structure 
factor from our numerical calculations has a finite width near its peak
position, which leads to a significant reduction on the value of $J$ 
as shown in Sec. \ref{scaling}.
So it is inaccurate to assume $|\k| = q$ for all $\k$. 
To improve it, we now discuss a more physically sound theory.

\section{Random Phase Approximation for STC} \lbl{rpa}

\subsection{Convective Current} \lbl{rpa_cc}

We now consider a different approach to STC (both PT and SDC) in RBC. 
For this purpose, we notice that although the instantaneous patterns in STC
are irregular and random in space 
\cite{mo_bo_93,as_st_93,xi_gu_93,be_fa_93,de_pe_94,xi_li_97}, 
some spatial symmetries can be restored if 
the system is averaged over a very long time. 
For example, for a laterally infinite system, 
it seems natural to assume that the time-averaged quantity
$\tv{\psi^2(\rr,t)}$ is uniform in space and that 
the two-point correlation function, defined as
$C(\rr_1,\rr_2) \equiv \tv{\psi(\rr_1,t) \psi(\rr_2,t)}/\tv{\psi^2(\rr_1,t)}$,
is translation invariant, i.e., $C(\rr_1,\rr_2) = C(\rr_1 - \rr_2)$. 
Then, one finds from Eq. \rf{cvcrnt} that $\tv{\psi^2(\rr,t)} = J$, and,  
from Eqs. \rf{ft} and \rf{dfs} that  
\bge
	\tv{\hpsi^*(\k_1,t)\hpsi(\k_2,t)} 
	= J \delta_{\k_1,\k_2} \hat{S}(\k_1).
	\lbl{tw_pt}
\nde
It is also easy to check that  
$\hat{S}(\k)$ is just the Fourier component of $C(\rr)$. 
 
	To understand the physical implications of the above result, 
we assume that
\bge
\hpsi(\k,t) = \hrho(\k,t) e^{i \hphi(\k,t)}, \lbl{am_ph}
\nde
with both $\hrho(\k,t)$ and
$\hphi(\k,t)$ real. Since $\hpsi^*(\k,t) = \hpsi(-\k,t)$, one has
$\hrho(-\k,t) = \hrho(\k,t)$, $\hphi(-\k,t) = - \hphi(\k,t)$
and $\hj(\k) = \tv{\hrho^2(\k,t)}$.
While both experimental measurements  and numerical 
calculations 
suggest that the amplitude has a well-defined 
time-averaged value with small fluctuations, the phase seems to have a rather
complicated, irregular time-dependence. 
If the amplitude and the phase can be assumed to be uncorrelated in time,  the
above result indicates that phases 
of different modes are totally {\it uncorrelated} in time. 
Since the $\delta_{\k_1,\k_2}$ factor in Eq. \rf{tw_pt} can be represented by 
\bge
	\delta_{\k_1,\k_2} = (2 \pi)^2 A^{-1} \delta(\k_1 - \k_2)  
	= \lim_{\xi_\phi \to +\infty} (2 \pi \xi^2_\phi/A)
	\exp[-\hf (\k_1 -\k_2)^2 \xi^2_\phi], \lbl{delta_gs}
\nde
we hence adapt
a {\it random phase approximation} (RPA) to STC in which 
\bge
	\tv{\exp[-i \hphi(\k_1,t) + i \hphi(\k_2,t)]} =
	\exp[-\hf (\k_1 -\k_2)^2 \xi^2_\phi]. 
	\lbl{rpa_two}
\nde
Here $\xi_\phi$ is a length determining the correlation between phases
of different modes.  We expect that 
$\xi_\phi = [A/2 \pi]^{1/2} \to + \infty$ for a laterally infinite system
but, as we will show,  this limit  should be taken only later on. 

Now, since the phases  are random in time, one may further 
approximate a four-phase correlation by products 
of  two-phase correlations, i.e.,  
\bgea
	&&\tv{\exp[-i \hphi(\k_1,t) - i \hphi(\k_2,t)
		+ i \hphi(\k_3,t) + i \hphi(\k_4,t)]}
	\nonumber \\
	&&\simeq \tv{\exp[-i \hphi(\k_1,t) - i \hphi(\k_2,t)]} \ 
	\tv{\exp[i \hphi(\k_3,t) + i \hphi(\k_4,t)]} 
	\nonumber \\
 	&& \quad +	\tv{\exp[-i \hphi(\k_1,t) + i \hphi(\k_3,t)]} \ 
	\tv{\exp[-i \hphi(\k_2,t) + i \hphi(\k_4,t)]} 
	\nonumber \\
	&& \quad +	\tv{\exp[-i \hphi(\k_1,t) + i \hphi(\k_4,t)]} \ 
	\tv{\exp[-i \hphi(\k_2,t) + i \hphi(\k_3,t)]}
	\nonumber \\
	&&= \exp\left[-\hf (\k_1 +\k_2)^2 \xi^2_\phi 
		- \hf (\k_3 +\k_4)^2 \xi^2_\phi\right]
	+ \exp\left[-\hf (\k_1 -\k_3)^2 \xi^2_\phi 
		-\hf (\k_2 -\k_4)^2 \xi^2_\phi\right] 
	\nonumber \\
	&& \quad + \exp\left[-\hf (\k_1 -\k_4)^2 \xi^2_\phi 
		-\hf (\k_2 -\k_3)^2 \xi^2_\phi\right].
	\lbl{rpa_four}
\ndea
Applying this and neglecting the correlations of $\hj(\k,t)$ such that 
$\tv{\hj(\k_1,t)\hj(\k_2,t)} \simeq \hj(\k_1) \hj(\k_2)$, one
may rewrite Eq. \rf{tveqcvcrnt} in a continuous Fourier space as
\bgea
	&&\pv{r(\epsilon,k)}_\k J
	=	\frac{A^4}{(2 \pi)^8} \int d \k_1 d \k_2 d \k_3 d \k_4 \,
	\frac{(2 \pi)^2}{A} \delta(\k_1 + \k_2 - \k_3 -\k_4)\,
	g(\hat{\k}_1 \cdot \hat{\k}_2) 
	\nonumber \\
	&&\hskip 6em	\times \left[\hj(\k_1) \hj(\k_3) 
	e^{-\hf (\k_1 +\k_2)^2 \xi^2_\phi -\hf (\k_3 +\k_4)^2 \xi^2_\phi}
	+ \hj(\k_1) \hj(\k_2) 
	e^{-\hf (\k_1 -\k_3)^2 \xi^2_\phi -\hf (\k_2 -\k_4)^2 \xi^2_\phi}
	\right.
	\nonumber \\
	&&\hskip 7.2em\left.	+\hj(\k_1) \hj(\k_2) 
	e^{-\hf (\k_1 -\k_4)^2 \xi^2_\phi -\hf (\k_2 -\k_3)^2 \xi^2_\phi}
	\right].
	\lbl{tveqcvcrnt_rpa}
\ndea
Since $\xi_\phi \to + \infty$, we may take
$g(\hat{\k}_1 \cdot \hat{\k}_2) =  g(-1)$ for the first term on the 
right hand side. Then, after some algebra, one finds that
\bge
	J = \frac{4 \pi \xi_\phi^2}{A} \frac{\pv{r(\epsilon,k)}_k}{g_\infty}
		= \frac{2 \pv{r(\epsilon,k)}_k}{g_\infty},
	\lbl{tvcvcrnt_rpa}
\nde
in addition to the conduction solution $J = 0$. 
Here $A = 2 \pi \xi_\phi^2$ has been used, 
$\hat{S}(\k)$ is assumed to be azimuthally uniform, and 
$g_\infty$ is defined in Eq. \rf{g_inf}. Notice that this result is valid for
both PT and SDC, but the exact values of $g_\infty$ and 
$\pv{r(\epsilon,k)}_k$  depends on the boundary conditions. 
While the value of $g_\infty$ for PT  has been given in Eq. \rf{g_pt}, 
the corresponding value for SDC is 
\bge
	g_{SDC} = 1.1319 + 0.0483 \sigma^{-1} + 0.0710 \sigma^{-2},
	\lbl{g_sdc}
\nde
which is obtained by interpolating the data for $g(\cos\alpha)$ given in
Ref. \cite{cr_80} for rigid-rigid boundaries 
and by integrating the consequent fitting function.

It is worthwhile to point out that if one takes 
the limit $\xi_\phi \to + \infty$ as early as in Eq. \rf{rpa_four}, one misses
the factor $2$ in Eq. \rf{tvcvcrnt_rpa}. 
To understand this, one should notice that Eq. \rf{rpa_four} reduces to
$\delta_{\k_1,-\k_2} \delta_{\k_3,-\k_4}+\delta_{\k_1,\k_3} \delta_{\k_2,\k_4}
+\delta_{\k_1,\k_4} \delta_{\k_2,\k_3}$ under the limit $\xi_\phi \to +\infty$
so that the constraint 
$\k_1 + \k_2 = \k_3 + \k_4$ in Eq. \rf{tveqcvcrnt}
is automatically satisfied. Consequently,
the delta function  
$(2 \pi)^2 A^{-1} \delta(\k_1 + \k_2 - \k_3 - \k_4)$ in Eq.
\rf{tveqcvcrnt_rpa}
is {\it not} necessary and can be replaced by
$1$. Considering that  this constraint is intrinsic in our problem
and physically needed, we believe that such a replacement is not justified.
In our calculation,
the limit $\xi_\phi \to + \infty$ is taken only at the last step of the 
calculation, 
which makes the constraint an important feature in Eq.~\rf{tveqcvcrnt_rpa}.
Our approach is also justified by our numerical results. 
It is obvious from Table \srf{table_pt} and \srf{table_sdc} 
that this factor of $2$ improves the 
agreement between our theory and our numerical results.

\subsection{Vorticity Current} \lbl{rpa_vc}

	We now use the RPA to calculate the time-averaged vorticity current. 
From Eqs. \rf{vtctcrnt} and \rf{rpa_four}, one finds that
\bgea
	&&\Omega
	=	\frac{A^4}{(2 \pi)^8} \int d \k_1 d \k_2 d \k_3 d \k_4 \,
	f(\k_1+\k_2;\k_2) f(\k_1+\k_2;\k_3)
	\frac{(2 \pi)^2}{A} \delta(\k_1 + \k_2 - \k_3 -\k_4)
	\nonumber \\
	&&\hskip 5em	\times \left[\hj(\k_1) \hj(\k_3) 
	e^{-\hf (\k_1 +\k_2)^2 \xi^2_\phi -\hf (\k_3 +\k_4)^2 \xi^2_\phi}
	+ \hj(\k_1) \hj(\k_2) 
	e^{-\hf (\k_1 -\k_3)^2 \xi^2_\phi -\hf (\k_2 -\k_4)^2 \xi^2_\phi}
	\right.
	\nonumber \\
	&&\hskip 7.2em\left.	+\hj(\k_1) \hj(\k_2) 
	e^{-\hf (\k_1 -\k_4)^2 \xi^2_\phi -\hf (\k_2 -\k_3)^2 \xi^2_\phi}
	\right],  \lbl{eq_tvvtctcrnt}
\ndea
where, from Eq. \rf{cplf}, 
\bgea
	&&f(\k_1+\k_2;\k_2) f(\k_1+\k_2;\k_3) \delta(\k_1+\k_2-\k_3-\k_4)
	\nonumber \\
	&&= \frac{g_m^2}{4 \sigma^2} \frac{(k_1^2 - k_2^2) (k_3^2 - k_4^2) 
		({\bb e}_z \cdot \k_1 \times \k_2)
		({\bb e}_z \cdot \k_3 \times \k_4)}
		{[|\k_1 + \k_2|^2 +c^2]\, [|\k_3 + \k_4|^2 +c^2]}
		\delta(\k_1+\k_2-\k_3-\k_4). \lbl{eq_ff}
\ndea
Notice that there is a singular point $\k_1 + \k_2 = 0$
in the above expression for free-free boundaries:
Since $c^2 = 0$ from Eq. \rf{para_fr_fr}, 
the value of the above expression depends on how the point is approached.
On the contrary, it is smooth everywhere for rigid-rigid boundaries, in which
case $c^2 = 10$ from Eq. \rf{para_rg_rg}.

The evaluation of $\Omega$ is rather complicated, which we present with
some details here.
We assume that $\hat{S}(\k)$ is azimuthally uniform, i.e., 
$\hj(\k) = 2 \pi A^{-1} J S(k)$ from Eqs. \rf{dfs}~-~\rf{ave}.
Clearly $\Omega$ consists of three terms. With $\q = \k_1 +\k_2$,
the first term, after some algebra, can be reduced to 
\bgea
	\Omega_1 = \frac{A}{(2 \pi)^4} \frac{g_m^2 J^2}{4 \sigma^2}
	&&\int d \q\,\frac{e^{-q^2 \xi_\phi^2}}{(q^2 + c^2)^2}
	\int d \k_1 \,S(k_1) (2 \q \cdot \k_1 -q^2) 
		({\bb e}_z \cdot \k_1 \times \q) \nonumber \\
	&&\times\int d \k_3 \,S(k_3) (2 \q \cdot \k_3 -q^2) 
		({\bb e}_z \cdot \k_3 \times \q).  \lbl{omega_1}
\ndea
It is easy to see that the angular integrals over $\k_1$ and $\k_3$ are
both zero, so $\Omega_1 = 0$. 
By interchanging $\k_3 \leftrightarrow \k_4$, one can show that 
the second and the third term  are identical, so 
$\Omega = 2 \Omega_2  = 2 \Omega_3$. With $\q = \k_1 - \k_3$, one finds
that
\bgea
	\Omega &=& \frac{A}{(2 \pi)^4} \frac{g_m^2 J^2}{2 \sigma^2}
	\int d \k_1 d \k_2 \, S(k_1) S(k_2) 
	\frac{(k_1^2 - k_2^2) ({\bb e}_z \cdot \k_1 \times \k_2)}
		{[|\k_1 + \k_2|^2 +c^2]^2}	\nonumber \\
	&& \times \int d \q\, e^{-q^2 \xi_\phi^2}
		[k_1^2 - k_2^2 -2 \q \cdot (\k_1 + \k_2)]\,
		[{\bb e}_z \cdot \k_1 \times \k_2
			-{\bb e}_z \cdot \q \times (\k_1 +\k_2)]
			\nonumber \\
	&=& \frac{g_m^2 J^2}{4 \sigma^2}
	\int dk_1\,k_1 S(k_1) \,\int d k_2 \, k_2 S(k_2) \,\Delta(k_1;k_2;c^2),
	\lbl{tvvtctcrnt}
\ndea
where we have used $A = 2 \pi \xi_\phi^2$ and 
\bgea
	\Delta(k_1;k_2;c^2) &=& (k_1^2 -k_2^2)^2 k_1^2 k_2^2 
	\int^\pi_0 \frac{d \alpha}{\pi} 
	\frac{\sin^2 \alpha}{[k_1^2 + k_2^2 + 2 k_1 k_2 \cos \alpha + c^2]^2}
	\nonumber \\
	&=& \frac{1}{4} (k_1^2 - k_2^2)^2 \left[
	\frac{k_1^2 + k_2^2 +c^2}{\sqrt{(k_1^2 - k_2^2)^2 + 2 c^2(k_1^2 +k_2^2)
		+c^4}} - 1 \right]. \lbl{Delta}
\ndea
For free-free boundaries, since $c^2 = 0$, one has
that
\bge
	\Delta(k_1;k_2;0) = \frac{1}{4} |k_1^2 - k_2^2|\, 
	[k_1^2 + k_2^2 - |k_1^2 - k_2^2|], \lbl{Delta_fr}
\nde
which has a second-order singularity at $k_1 = k_2$ and is due to the 
singularity in Eq. \rf{eq_ff}. In comparison,
the function $\Delta(k_1;k_2;c^2)$ 
is analytic everywhere for rigid-rigid
boundaries with $c^2 = 10$. While
$\Delta(k_1;k_2;0) \sim |k_1^2 - k_2^2|$ for free-free boundaries,
correspondingly $\Delta(k_1;k_2;c^2) \sim (k_1^2 -k_2^2)^2$ for rigid-rigid
boundaries. As we will show in the next section, 
this has a significant consequence to the
properties of STC.

\section{Scaling Relations in STC} \lbl{scaling}

\subsection{General} \lbl{scaling_gen}

To evaluate the convective current $J$ and the vorticity current
$\Omega$, 
one must know the structure factor $S(k)$ which, however, is beyond our
present theory. We thus  turn to phenomenological arguments. 
We define a two-point correlation length as 
\bge
	\xi_2 = \left[\pv{k^2}_k - \pv{k}_k^2\right]^{-1/2} \lbl{lngth}.
\nde
Then we 
assume that the structure factor satisfies the following {\it scaling} form
\bge
	k S(k) = \xi_{2} \calf[(k - k_{max}) \xi_{2}],  \lbl{s_scaling}
\nde
where $\km$ is the peak position
of $k S(k)$ and  $\calf(x)$ is the scaling function satisfying 
$\int_{-\infty}^\infty d x\, \calf(x) = 1$.  
[Since $k \ge 0$ in $k S(k)$,
the lower limit for $\calf(x)$ is $- k_{max} \xi_2$, which we approximate
by $-\infty$.]
Inserting $k = \km + x \xi_2^{-1}$ 
and Eq. \rf{s_scaling} into Eq. \rf{lngth},
one gets that $\pv{x^2}_x - \pv{x}_x^2 = 1$, where we have used the notation
\bge
	\pv{F(x)}_{x} 
	=\int_{-\infty}^\infty d x \, \calf(x) F(x).
		\lbl{x_ave}
\nde
It is also easy to see that $\pv{k}_k = \km + \xit^{-1} \pv{x}_x$. 

For very large $\xit$, one may take $k = \km + x \xi_2^{-1}$ in
Eqs. \rf{tvcvcrnt_rpa} and \rf{tvvtctcrnt} and    
expand the results in order of $1/\xi_2$. 
It is easy to find from Eq. \rf{r} that
\bge
	\pv{r(\epsilon,k)}_k = r(\epsilon,\km) 
		- b (b^2 - 1) \pv{x}_x k_c \xi_0^2/\xit
		- \hf (3 b^2 - 1) \pv{x^2}_x \xi_0^2/\xi_2^2 
		+ {\cal O}(1/\xit^3),
		\lbl{apprx_r}
\nde
where $ b = \km/k_c$. We expect 
that $b \simeq 1 + b_1 \epsilon$
in STC. So for small enough $\epsilon$, one has  that
\bge
	\pv{r(\epsilon,k)}_k \approx 
	\epsilon - \pv{x^2}_x \xi_0^2/\xi_2^2, \lbl{asmp_r}
\nde
and, from Eq. \rf{tvcvcrnt_rpa}, that
\bge 
	J \approx \frac{2}{g_\infty}
	\left[\epsilon - \pv{x^2}_x\frac{\xi_0^2}{\xi_2^2}\right], 
	\lbl{tvcvcrnt_rpa_asmp}
\nde
with $g_\infty = g_{PT}$ or $g_\infty = g_{SDC}$ depending on the boundary
conditions.
This expression depends on two unknown but experimentally measurable
quantities $\xit$ and $\pv{x^2}_x = 1 + \pv{x}_x^2$. If ${\cal F}(x)$ is 
symmetric, then $\pv{x}_x = 0$ and $\pv{x^2}_x = 1$. In general, however, one
has $\pv{x^2}_x \ge 1$.  
Notice also that although
mean-flow couplings are not explicitly present in Eq. \rf{tveqcvcrnt}, they
affect the value of the convective current via the structure
factor $\hat{S}(\k)$ [see Eq. \rf{keqcvcrnt}]
and the two-point correlation  length $\xi_2$.  
For PT, since 
$\xit \simeq \frac{3}{2} \xi_0/\epsilon^\hf$ \cite{xi_li_97}, the contribution
from $\xit$ reduces the value of $J$ quite significantly. 
This feature is absent in the
simple theory presented in Sec. \srf{smpl_th}. 

One may evaluate the vorticity current in the same way. From
Eqs. \rf{tvvtctcrnt} and \rf{Delta_fr}, one gets for free-free boundaries that
\bge
	\Omega \approx 
	\frac{g_m^2 k_c^3}{4 \sigma^2}\frac{J^2}{\xi_2}
	\pv{|x_1-x_2|}_{x_1,x_2},
	\lbl{Omega_scaling_fr_fr}
\nde
where we have used $\km \approx k_c$.
The quantity $\pv{|x_1-x_2|}_{x_1,x_2}$ is related to the width of the scaling
function ${\cal F}(x)$ and, since  
$\pv{|x_1-x_2|}_{x_1,x_2} \le \sqrt{\pv{(x_1-x_2)^2}_{x_1,x_2}} = \sqrt{2}$,
we expect $\pv{|x_1-x_2|}_{x_1,x_2}$ to be of order of unity.
Similarly, we find from Eqs. \rf{tvvtctcrnt} and \rf{Delta} for rigid-rigid boundaries that
\bge
	\Omega \approx 
	\frac{g_m^2 k_c^2}{2 \sigma^2}\frac{J^2}{\xi_2^2}
	\left[\frac{2 k_c^2 + c^2}{\sqrt{4 k_c^2 c^2 + c^4}} - 1\right].
	\lbl{Omega_scaling_rg_rg}
\nde
Clearly, by 
phenomenological arguments,
we can express $J$ and $\Omega$ in STC in terms of measurable
quantities.

\subsection{PT} \lbl{scaling_pt}

For PT, we further assume  power law behaviors for the two-point correlation 
length, the convective current and the vorticity current 
such  as
\bge
	\xi_2 \approx \xi_{2,0} \te^{-\nu}, \quad
	J \approx J_0 \epsilon^\mu \quad {\rm and} \quad 
	\Omega \approx  \Omega_0 \epsilon^\lambda. \lbl{scaling_exponts}
\nde
Then, from Eq. \rf{Omega_scaling_fr_fr}, we find 
the following scaling relation 
\bge
		\lambda = 2 \mu + \nu.  
		\lbl{scaling_fr_fr}
\nde
Recalling Eq. \rf{tvcvcrnt_rpa_asmp}, 
one obtains that
\bge
	J \approx \frac{2}{g_{PT}}
	\left[\epsilon -\pv{x^2}_x
	\frac{\xi_0^2}{\xi_{2,0}^2} \epsilon^{2 \nu}\right]
	\approx J_0 \epsilon^{\mu}. \lbl{J_scaling}
\nde
Since $J$ is positive by definition, the values of the 
exponents  satisfy 
\bge
	\mu = 1, \quad \nu \ge 1/2 \quad
	{\rm and} \quad \lambda = 2 + \nu \ge 5/2 \quad \quad {\rm in \ PT}.
		\lbl{exponents_fr}
\nde
It is very likely that $\nu = 1/2$, hence, $\lambda = 5/2$. If so, 
then one finds from Eqs. \rf{tvcvcrnt_rpa_asmp} and \rf{Omega_scaling_fr_fr}
that 
\bge
	J_0 = \frac{2}{g_{PT}}
	\left[1 - \pv{x^2}_x \frac{\xi_0^2}{\xi_{2,0}^2}\right]
	\quad {\rm and} \quad 
	\Omega_0 = 
	\frac{g_m^2 k_c^3 J_0^2}{4 \sigma^2 \xi_{2,0}}
	\pv{|x_1-x_2|}_{x_1,x_2},
	\lbl{amps_fr_fr}
\nde
which depend on three phenomenological parameters $\xi_{2,0}$, $\pv{x^2}_x$ 
and $\pv{|x_1-x_2|}_{x_1,x_2}$. If $\nu > 1/2$, then $J_0 = 2/g_{PT}$
since the $\epsilon^{2 \nu}$ term
in Eq. \rf{J_scaling} contributes only to the leading correction to scaling. 
It is interesting to notice that the amplitude equations coupled 
with mean-flow \cite{si_zi_81} predicts for free-free boundaries
that $\Omega \sim \epsilon^{5/2}$ for almost perfect parallel rolls. 

We now verify our predictions for the power laws in PT
by our numerical solutions. We have carried out 
large-scale numerical calculations of the three-dimensional Boussinesq 
equations  under free-free boundaries for fluids of $\sigma = 0.5$ 
\cite{xi_li_97}. We have confirmed in Ref. \cite{xi_li_97}
that the structure factor in PT satisfies
the scaling form \rf{s_scaling}. 
From Table \ref{table_pt}, one can see that our theoretical 
and our numerical results
are in very good agreement for the exponents. The scaling relation
Eq. \rf{scaling_fr_fr} is confirmed within our numerical uncertainties.   
The comparison between the corresponding amplitudes, however, is only 
moderately successful. Calculations of $\xi_{2,0}$, $\pv{x^2}_x$ and
$\pv{|x_1 - x_2|}_{x_1,x_2}$ 
are obviously beyond the present theory, so we take
our numerical result for $\xi_{2,0}$. 
Since our numerical results for $\pv{x^2}_x$ and $\pv{|x_1 - x_2|}_{x_1,x_2}$
are too sensitive to the large value cutoff to be meaningful, see discussions
in the next section, we assume equalities in 
$\pv{x^2}_x = 1 + \pv{x}_x^2 \ge 1$ and 
$\pv{|x_1 - x_2|}_{x_1,x_2} \le \sqrt{\pv{(x_1-x_2)^2}_{x_1,x_2}} = \sqrt{2}$.
 From Eqs. \rf{amps_fr_fr}, \rf{para_fr_fr} and \rf{g_pt}, one gets 
$J_0 \simeq 0.972$, which is about $20 \%$ larger than the numerical value.
A non-zero value of $\pv{x}_x$ will apparently reduce the theoretical
value of $J_0$ in the right direction.  
It is worthwhile to point out that, since $\xi_{2,0} \simeq (3/2) \xi_0$ 
\cite{xi_li_97},
the value of $J_0$ is reduced significantly owing to the finite width of the
power spectrum. On the other hand, one finds that
$\Omega_0 \simeq 454.7 \pv{|x_1-x_2|}_{x_1,x_2} = 643.0$ as an upper bound,  
which is about ten times larger
than our numerical result. Nevertheless,  
we note that while our theory is based on the
two-dimensional GSH equations, our numerical calculations are done for
the three-dimensional Boussinesq equations. Although the former is very good
in reproducing qualitative features of RBC, it
may not be quantitatively accurate in modeling RBC \cite{cr_ho_93,de_pe_94}. 
So one should be cautious in comparing
the results from the GSH equations 
with those from real experiments 
or those from numerical calculations with hydrodynamical equations. 

\subsection{SDC} \lbl{scaling_sdc}

The situation for SDC, however, is more subtle since the roll-to-SDC
transition occurs at a positive temperature $\epsilon_T$ 
\cite{mo_bo_93,as_st_93,xi_gu_93,be_fa_93,de_pe_94}. Consequently, 
several competing scaling scenarios are possible in SDC, the choice of which  
depends on the character of
the transition. A more thorough examination on the issue will
be presented elsewhere \cite{xi_li_97_2}. 
We mention that the same power laws as Eq. \rf{scaling_exponts} can be 
defined for SDC. But
instead of Eq. \rf{scaling_fr_fr}, one finds from Eqs. \rf{Omega_scaling_rg_rg}
and \rf{scaling_exponts} the following scaling relation 
\bge
        \lambda = 2 \mu + 2 \nu. 
                \lbl{scaling_rg_rg}
\nde
By the same arguments leading to Eq. \rf{exponents_fr},
one gets that 
\bge
	\mu = 1, \quad \nu \ge 1/2 \quad
	{\rm and} \quad \lambda = 2 + 2 \nu \ge 3 \quad \quad {\rm in \ SDC}.
		\lbl{exponents_rg}
\nde
The conclusion that different scaling relations hold for PT and SDC 
can be traced back to Eqs. \rf{Delta} and \rf{Delta_fr} 
via the different behaviors of
$\Delta(k_1;k_2;c^2)$ at $k_1 = k_2$. 

In order to test our theory of SDC, 
we have carried out systematic numerical studies of SDC with the GSH equations 
\cite{xi_li_97_2}.
For simplicity, we take $g(\cos \alpha) = g$ as a constant so, 
from Eq. \rf{g_inf}, $g_\infty = 3 g$.  For numerical
convenience, following Refs. \cite{xi_gu_93,xi_vi_92},
we rescale the GSH equations such as
\bge
\begin{tabular}{llll}
	$\rr \to k_c^{-1} \rr'$,& $t \to (4 \tau_0/k_c^2 \xi_0^2) t'$,
		\quad\quad&
	$\psi \to (k_c \xi_0/2\sqrt{g}) \psi'$, \quad \quad & 
	$\zeta \to (g_m \tau_0 k_c^2/g) \zeta'$, \\
	$\epsilon \to (k_c^2 \xi_0^2/4) \epsilon'$, \quad \quad&
	$\sigma \to (\xi_0^2/4 \tau_0) \sigma'$, & $c^2 \to k_c^2 c'^2$, &
	$g_m \to (g \xi_0^2/4 \tau_0^2 k_c^2) g'_m$, 
\end{tabular}
	\lbl{rescale}
\nde
which leads to the rescaled GSH equations
\bgea
         \partial_{t'} \psi' + g'_m \bb U' \cdot \bb \nabla' \psi'
        &=& \left[\epsilon' - ({\bb \nabla'}^2 + 1)^2 \right] \psi' - \psi'^3, 
		\lbl{rescale_gsh} \\
        \left[\partial_{t'}-\sigma' ({\bb \nabla'}^2 - c'^2)\right] 
		{\bb \nabla'}^2 \zeta'
        &=& {\bb e}_z \cdot \left[{\bb \nabla'}( {\bb \nabla'}^2 \psi')
		\times {\bb \nabla'} \psi'\right], 
		\lbl{rescale_mf}
\ndea
where $\bb U'(\rr') =  {\bb \nabla'} \zeta'(\rr',t') \times {\bb e}_z$.
Now the time-averaged convective current \rf{tvcvcrnt_rpa_asmp} and the
time-averaged vorticity current \rf{Omega_scaling_rg_rg}  are rescaled into
\bge
	J'_{SDC} \approx 
	\frac{2}{3}\left[\epsilon'  - \frac{4 \pv{x^2}_x}{\xi'^2_2}\right], 
	\lbl{tvcvcrnt_rescale}
\nde 
and, 
\bge
	\Omega'_{SDC} \approx \frac{1}{2 \sigma'^2}
	\left[\frac{2 + c'^2}{\sqrt{4 c'^2 + c'^4}} -1\right] 
	\frac{J'^2}{\xit'^2}. 
	\lbl{tvvtctcrnt_rescale}
\nde
From Eq. \rf{para_rg_rg}, one finds that 
$\epsilon = 0.3594 \epsilon'$ for rigid-rigid boundaries. 
In principle, for a given 
$\sigma$ and a suitably chosen $g$, the parameters $g'_m$, $\sigma'$ and
$c'^2$ are determined by Eqs. \rf{para_rg_rg} and \rf{rescale}. 
Again following Ref. \cite{xi_gu_93}, 
we simply choose $g'_m=50$, $\sigma'=1.0$ and $c'^{2}=2.0$.
Details of the numerical studies of SDC are presented elsewhere 
\cite{xi_li_97_2}.

One crucial assumption in our theory of SDC is that
the structure factor $S(k)$ has a scaling form like Eq. \rf{s_scaling}.
So it is very important to verify this assumption. 
In the insert of Fig. \ref{fig_scaling_sdc}, the results for $k' S'(k')$ for 
$\epsilon' = 0.55$, $0.65$ and $0.8$, corresponding to  SDC states, 
are plotted. [The structure factor $S'(k')$ is nomalized by 
$\int^\infty_0 d k'\,k' S'(k') = 1$.]
To check whether a scaling form like
Eq. \rf{s_scaling} holds, we take
the two-point correlation length $\xit'$ 
from our numerical results  and choose $k'_{max}$ to give the best fit to 
scaling. 
For each $\epsilon'$  within $0.55 \le \epsilon' \le 0.8$,
we hence find a corresponding function $\calf(x)$  of SDC, 
which is shown  
in Fig. \ref{fig_scaling_sdc}. 
As one can see, all the data collapse into one
single curve. The scattering of the data near $\km'$ is due to our numerical
uncertainties and is within the corresponding error bars. 
So the existence of a scaling form of
$k S(k)$ is verified within our numerical uncertainties for SDC. 

We now compare our numerical results for $J'$ and $\Omega'$
with those from our theory. Theoretical results are presented in 
Eqs. \rf{tvcvcrnt_rescale} and \rf{tvvtctcrnt_rescale}.
We fit our numerical data with power laws such as
$\xit' = \xi'_{2,0} (\epsilon' - \epsilon_c')^{-\nu}$, 
$J'=J_0'(\epsilon'-\epsilon_c')^\mu - J_\xi' \xit'^{-2}$ and 
$\Omega'=\Omega_0' (\epsilon'-\epsilon_c')^\lambda$ 
with $\epsilon_c' = 0.002$, see Ref. \cite{xi_li_97_2}.
The non-zero value of $\epsilon_c'$ is likely due to finite-size
effects. 
In Table \ref{table_sdc},
we summarize both theoretical and numerical
results for $J'$, $\Omega'$ and $\xi'$ for SDC. 
We actually 
put $\mu = 1$ in our fitting of $J'$, so the agreement with this 
is trivial. The inequality for the theoretical value of 
$J'_\xi$ is from $\pv{x^2}_x \ge 1$. Since
the  calculations of  $\xi'_{2,0}$ and $\nu$ are beyond our theory, 
we use the corresponding
numerical results in calculating $\Omega_0'$ and $\lambda$. 
Clearly the scaling relation Eq. \rf{scaling_rg_rg} is approximately verified.
The prediction for $J'$ is very good. The prediction for
the value of $\Omega_0'$, however, is larger than the corresponding numerical
result by a few magnitudes. The cause for such a big discrepancy, at present,
is not clear to us. Considering that $\omega_z(\rr,t)$ has a highly 
localized structure in real space \cite{xi_gu_95,xi_li_97_2}, 
it is possible that our 
numerical calculation is not long enough to sample all the phase space.
It is also possible that assumptions in our theory are not sufficient to
describe the behavior of $\Omega$. 
In comparison with the situation in PT, which is discussed in 
Sec. \srf{scaling}(B), the success of our theory in describing SDC is 
not as satisfactory. 
Further improvement of it is obviously valuable. 

\section{Is Porod's law valid?} \lbl{porod} 

In phase ordering, a 
sharp interface exists between domains of different phases. 
Consequently, the real-space correlation function $C(r)$
is proportional to $r/L$ 
at short distances, where $L$ is a characteristic length of the system 
\cite{bray}. Then the corresponding 
structure factor, which is the Fourier transformation of $C(r)$, 
behaves like $S(k) \sim 1/L k^3$ for large $k$ in two-dimensional space. 
This large $k$ behavior of $S(k)$ is known as Porod's law \cite{po_82,bray}. 
It is easy to check that 
the two-point correlation length defined in Eq. \rf{lngth} is very
sensitive to the large $k$ cutoff if Porod's law is valid. As a result, 
other criteria are needed to
define  a better behaved characteristic length, say $L$, of the system.

For the convective patterns in RBC, smooth interfaces are always present
between hot, rising fluid and cold, sinking fluid. In the ordered states,
the patterns can be described by a few sine or cosine modes. Correspondingly,
the structure factor consists of only several sharp peaks. Porod's law is not
relevant in this case. 
But in STC, an infinite number of modes are excited, including those
large $k$ modes.
Then, a natural question can be raised: Is Porod's law valid in STC? 
Considering that the shape of the
interface between different domains appears random and the motion of it seems
chaotic, an intuitive argument is rather difficult. 
In this section, we present our efforts in this direction. 

To start, we take the scaling form Eq. \rf{s_scaling}  of $kS(k)$ but replace
$\xi_2$ with a characteristic length $\xis$ in case $\xi_2$ is cut-off
dependent. If $\xi_2$ is well-defined, from Eqs. \rf{lngth} and 
\rf{s_scaling} (with $\xi_2$ replaced by $\xis$), one gets that
\bge
	\xis \approx \xi_2 \sqrt{\pv{x^2}_x - \pv{x}_x^2}. \lbl{xi_xi}
\nde
So $\xis$ and $\xi_2$ are identical up to an overall constant.
Recall the following formulas
\bge
	e^{i \k \cdot \rr} = J_0(k r) + 2 \sum_{m=1}^\infty i^m 
		J_m(k r) \cos m \alpha, \lbl{bssl_1}
\nde
\bge
	J_0(x+y) = J_0(x) J_0(y) + 2 \sum_{m=1}^\infty (-1)^m J_m(x) J_m(y),
		\lbl{bssl_2}
\nde
where $\alpha$ is the angle between $\k$ and $\rr$, 
and $J_m(x)$ is the $m$-th order Bessel function. It is straightforward 
to show that
\bge
	C(r) = \int d k \, k S(k) J_0(k r) 
	= J_0(\km r) C_0(r/\xis) 
	+ 2 \sum_{m=1}^\infty (-1)^m J_m(\km r) C_m(r/\xis), \lbl{c_r}
\nde
where
\bge
	C_m(r/\xis) = \int^{+\infty}_{-\infty} d x\,\calf(x) J_m(x r/\xis).
		\lbl{c_m}
\nde

Since $J_m(y) \sim y^m$, one has that $C_m(y)/y^i \to 0$ as $y \to 0^+$ for all
$i = 0, 1, \cdots, m-1$. So, neglecting the possible presence of singularity,
we assume the following expansions
\bge
	C_m(y) = y^m \sum_{i=0}^\infty C_{mi} \,y^i 
	\quad {\rm for\ small} \ y>0.
	\lbl{c_expd}
\nde
[One cannot apply the small $y$ expansion of $J_m(y)$ in Eq. 
\rf{c_m} since, for any fixed $r/\xis$, the integral is dominated by
those $x$'s such that $x r/\xis$ is not small.]
Since $J_0(y) \to 1$ as $y \to 0^+$, one finds that 
$C_{00} = \int^{+\infty}_{-\infty} d x\, \calf(x) = 1$. Now it is easy to see
from Eq. \rf{c_r} that 
\bge
	C(r) \approx 1 + C_{01} r/\xis \quad
	{\rm for} \quad  \km r \simeq k_c r \ll 1. \lbl{c_r_small}
\nde
While the constant term contributes an unmeasurable $\delta(\k)$ to 
$\hat{S}(\k)$, the linear term leads to the Porod's law, i.e.,
\bge
	S(k) \sim 1/\xis k^3 \quad {\rm for} \quad k \gg k_c.
\nde
It is worthwhile to mention that the $1/\xis$ dependence is as important as
the $1/k^3$ dependence \cite{po_82,bray}. In phase ordering, a large $k$
cutoff exists so that Porod's law is valid for those
$k$'s smaller than this cutoff \cite{bray}. It is not clear whether such
a large cutoff exists in STC. One possibility is that this
cutoff exists and is of the same order of $k_c$, in which case 
Porod's law is limited to
such a narrow range in $k$ space that verification of it is almost impossible.

Assuming $\xit$ is well-defined, 
we plot $\xi_2 k^3 S(k)$ vs. $k$ for both PT and 
SDC in Fig. \ref{fig_porod}. 
The data for PT are obtained from our numerical solutions of the 
three-dimensional Boussinesq equations \cite{xi_li_97}, evaluated at the 
mid-plane. The data for SDC are from our numerical calculations of the GSH
model \cite{xi_li_97_2}.  
As one can see, the value of $\xi_2 k^3 S(k)$ in PT seems to 
approach a constant at large $k$, insensitive to the exact value of 
$\epsilon$. So Porod's law might 
be valid in PT.  But the value of  $\xit k^3 S(k)$ seems 
to {\it increase} for large $k$ in SDC!
However, it is known the GSH model introduces an artificial short-ranged (hence
large $k$) cross-roll instability \cite{gr_cr_85}, so the large $k$ behavior
in the GSH model might be different from those in real systems. 
Furthermore, owing to the finite grid size used
in our numerics, we are not sure how numerical noise might affect
the large $k$ behavior in both PT and SDC.
For this reason, we believe that
more accurate data are needed for a definite conclusion.    
Even so, one sees immediately how sensitive
the two-point correlation length 
$\xi_2$ defined by Eq. \rf{lngth} could be to the 
large $k$ cutoff.  
So it is useful to define a less sensitive characteristic length, 
say $\xis$, of the system. One obvious choice is the inverse of
the full width at the half peak (FWHP) of $k S(k)$.
Since one can easily find 
a function $\calf(x)$ satisfying Eq. \rf{s_scaling} for each $\xis$  
(replacing $\xit$ with $\xis$), 
this provides the easiest way to check whether a scaling form exists. 
If the system is inside the scaling range,
all $\calf(x)$'s so defined should collapse into a single curve. 
One must, of course, normalize $S(k)$ by $\int d k\, kS(k) = 1$ first. 
But this normalization 
is much less sensitive to the large $k$ cutoff than $\xit$ is.  
As shown in Eq. \rf{xi_xi}, if $\xit$ is well defined, $\xis$ and $\xit$
are simply proportional to each other inside the scaling range. This
is not true if the system is outside the scaling range.

\section{Discussion} \lbl{disc}

Our phenomenological theory for STC in RBC 
depends on two basic assumptions. In Sec. \srf{rpa}, we assume that the time-averaged 
two-point correlation function is translation invariant in real space and we hence adapt a random
phase approximation to STC. In Sec. \srf{scaling}, we further assume that 
the structure factor 
satisfies a scaling form such as 
$k S(k) = \xit {\cal F}[(k- \km)\xit]$. In comparison with
similar scaling forms in critical phenomena, critical dynamics and
phase ordering \cite{bray,ma_94}, we find it necessary to replace $k$ with
$k - \km$ in the scaling form. The physical origin of 
this replacement is due to  the fact that patterns in RBC have an intrinsic
wavenumber, which is close to $k_c$. By the same reason, we find it necessary
to seek the scaling form of $k S(k)$ instead of $S(k)$, where the $k$
factor comes from $d \k = k\, dk\, d \alpha$ in two-dimensional $k$-space.
The existence of the scaling forms in critical phenomena and critical dynamics
is rooted in the scaling
invariance of long wavelength fluctuations in the system
and is associated, respectively, with a 
fixed point in renormalization group theory \cite{ma_94}. Its physical
origin in STC is yet unknown.
In Sec. \srf{scaling}, we have confirmed the scaling form
of $S(k)$ within our numerical accuracy.   
Since $k \ge 0$ in 
$k S(k)$, the lower limit for the scaling function ${\cal F}(x)$ is 
$-\km \xit$, which is $\epsilon$ dependent. So the violation of scaling is 
almost certain for very small $k$. We cannot rule out from our 
numerical data that this 
scaling form might also be violated for very large $k$. It is not clear
currently in what range the scaling form is valid. 

As we discussed in Sec. \srf{porod}, the two-point correlation length $\xit$
is cutoff dependent if Porod's law is valid for STC in RBC. In principle, 
there is another
disadvantage to choose $\xit$ as a characteristic length. It is easy to see
from Eq. \rf{s_scaling} that $\pv{k}_k = \km + \xit^{-1} \pv{x}_x$, so
$\pv{k}_k$ is shifted from $\km$ by $\pv{x}_x/\xit$. Because of this, an 
unknown parameter $\pv{x^2}_x$ is introduced in Eq. \rf{tvcvcrnt_rpa_asmp}.
This $\pv{x^2}_x$ parameter can be easily removed by defining a new length
$\xi_s = [\pv{(k - \km)^2}_k]^{-1/2}$, instead of Eq. \rf{lngth}. Then
one simply has $\pv{x^2}_x = 1$ if $\xit$ is replaced by $\xi_s$ in
Eq. \rf{s_scaling}. In practice, however, our numerical data are not
accurate enough to determine $\km$ precisely. Consequently, there is no
practical 
advantage for us to use $\xi_s$ instead of $\xit$. This may not be
true for experimentalists since their data are much more accurate. 
Of course, it is also to be tested whether the structure factor can satisfy
a scaling form like Eq. \rf{s_scaling} with respect to $\xi_s$ so defined. 

In summary, we present a phenomenological theory for STC in RBC. We calculate
analytically the time-averaged convective current $J$ and the time-averaged
vorticity current $\Omega$ in both PT and SDC
as functions of $\epsilon$ and $\xit$. Our theory
is successful for both PT and SDC, despite the need 
for a better quantitative 
result for $\Omega$ in SDC. 
We believe that our theoretical results
will be useful in understanding the complicated behavior of STC in RBC. 
We also believe that our theory provides a new 
approach to STC and
also raises some interesting questions. 
For example, how can one 
calculate the structure factor $S(k)$ and the two-point
correlation length $\xit$ analytically? Is it possible that certain global 
quantities in STC form a complete set in the same way as temperature,
pressure and density do for thermodynamic systems?  Can we derive
some effective variational principle in terms of global quantities?
How far can we apply the ideas in critical phenomena to study STC? 
Since our assumptions are quite general, it will also be interesting to see
whether our theory can be generalized to STC in other systems 
\cite{cr_ho_93,gr_96}.

\begin{center} Acknowledgment \end{center}

	X.J.L and J.D.G are supported by the National Science Foundation 
under Grant No. DMR-9596202. H.W.X. is supported by Research Corporation 
under Grant No. CC4250. 
Numerical work reported
here are carried out on the Cray-C90 at the Pittsburgh
Supercomputing  Center and Cray-YMP8 at the Ohio Supercomputer Center.

%

\begin{table}
\vskip 0.5in
\caption{Time-averaged convective current $J \approx J_0 \epsilon^\mu$, 
time-averaged vorticity current $\Omega \approx \Omega_0 \epsilon^\lambda$ 
and two-point correlation length $\xi_2 \approx \xi_{2,0} \epsilon^{-\nu}$ 
in PT with $\sigma = 0.5$. For theoretical result of $\nu$, 
we assume equality in Eq. (54). See also discussions in Sec. V(B).}
\lbl{table_pt}
\begin{tabular}{lcccccc}
 & $\mu$ & $\nu$ & $\lambda$ & $\xi_{2,0}$ & $J_0$ & $\Omega_0$ \\ \hline
Numerics & $1.034 \pm 0.025$ & $0.472 \pm 0.016$ & $2.55 \pm 0.10$ 
	& $0.82 \pm 0.04$ & $0.787 \pm 0.019$ & $70.1 \pm 1.0$ \\ 
Theory & $1$ & $1/2$ & $5/2$ & ---  
	& $0.972$ & $643.0$ \\
\end{tabular}
\end{table}


\begin{table}
\caption{Time-averaged convective current 
$J' \approx J'_0 \epsilon'^\mu  - J'_\xi \xit'^{-2}$, 
time-averaged vorticity current $\Omega' \approx \Omega'_0 \epsilon'^\lambda$ 
and two-point correlation length $\xi_2' \approx \xi'_{2,0} \epsilon'^{-\nu}$ 
in SDC, with $g'_m= 50$, $\sigma' = 1.0$ and $c'^2 = 2.0$. For numerical 
results,
we actually use $\epsilon' - \epsilon_c'$ with $\epsilon_c' = 0.002$ 
instead of $\epsilon'$ for data fittings. For more details, see Ref. [26].}
\lbl{table_sdc}
\begin{tabular}{lccccccc}
 & $\mu$ & $\nu$ & $\lambda$ & $\xi'_{2,0}$ & $J'_0$ & $J'_\xi$ & $\Omega'_0$ \\ 
	\hline
Numerics & $1$ & $0.72 \pm 0.05$ & $3.0 \pm 0.1$ & $6.8 \pm 0.2$ 
	& $0.64 \pm 0.02$ & $2.9 \pm 0.9$ & $(3.0 \pm 0.2) \times 10^{-8}$ \\ 
Theory & $1$ & $\ge 1/2$ & $3.4 \pm 0.1$ & --- & $2/3$
	& $\ge 8/3$ & $3.7 \times 10^{-4}$ \\
\end{tabular}
\end{table}

\newpage
\centerline{\large FIGURE \ CAPTIONS}
\vskip 0.5in

	Figure \ref{fig_ring}. Allowed configurations of wavenumbers satisfying $k_1 = k_2 = k_3 = k_4$
and $\k_1+\k_2=\k_3+\k_4$: (a)$\k_1+\k_2 = \k_3+\k_4 =0$; (b)$\k_1 \neq \k_2$ and $\k_1 + \k_2 \neq 0$;
or (c)$\k_1=\k_2=\k_3=\k_4$. 

	Figure \ref{fig_scaling_sdc}. A plot of $k' S'(k')/\xit'$ 
vs. $x = (k' - \km') \xit'$ for $0.55 \le \epsilon' \le 0.8$ in SDC, 
showing scaling and the scaling function
${\cal F}(x)$ defined in the text. The scattering of the data is within our 
numerical uncertainties. 
Insert: The time-averaged function
$k' S'(k')$ vs. $k'$ for $\epsilon' = 0.55$, $0.65$ and $0.8$ in SDC. 

	Figure \ref{fig_porod}. Plots of $\xit k^3 S(k)$ vs. $k$ in (a) PT
and (b) SDC. The error bars are plotted only for (a) $\epsilon = 0.05$ in PT
and (b) $\epsilon' = 0.65$ in SDC.

\newpage
\begin{figure}
\vskip 0.5in
\epsfxsize = 5.5in
\epsfysize = 5.5in
\centering
\epsfbox{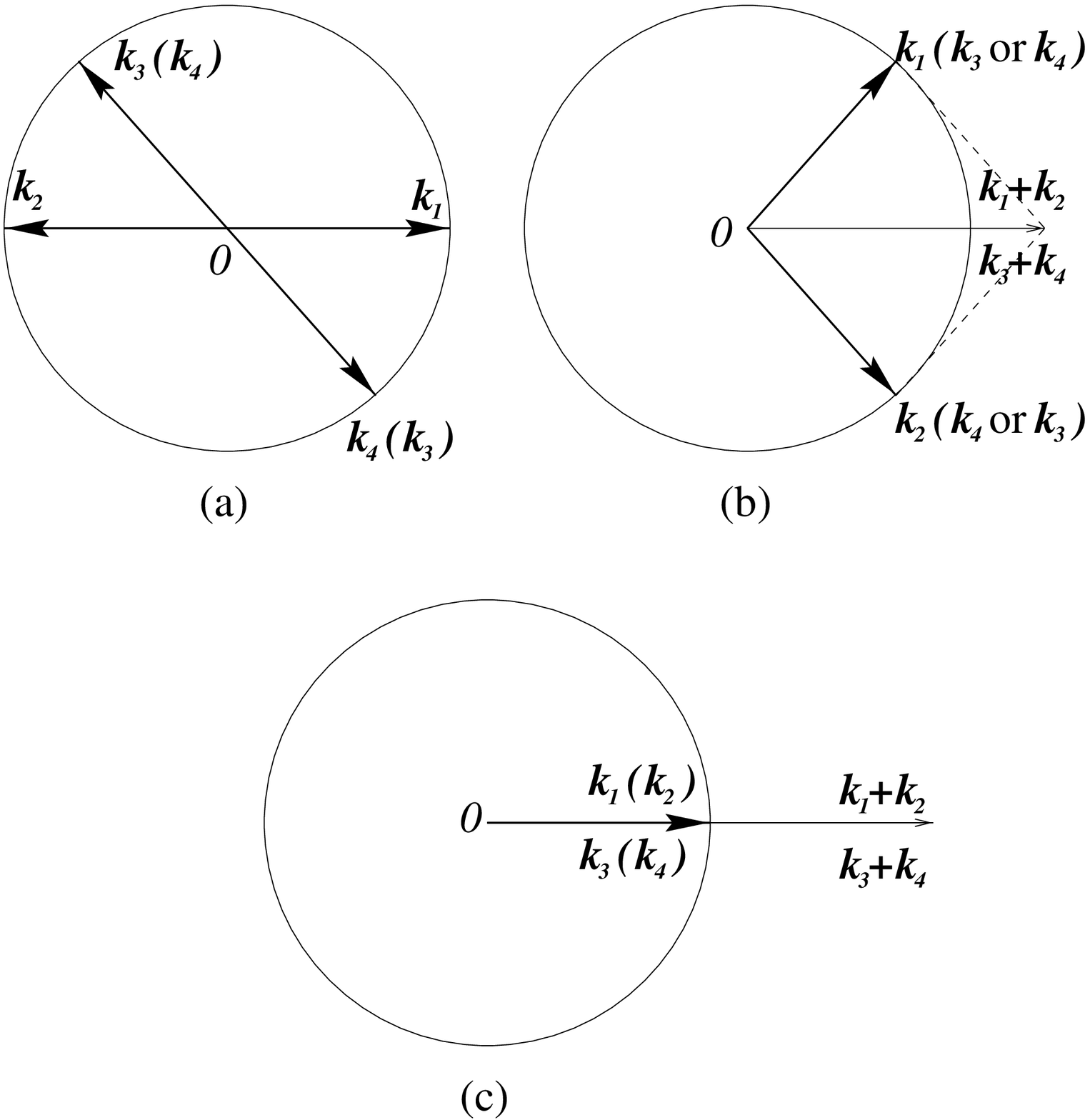}
\caption{}
\lbl{fig_ring}
\end{figure}

\newpage
\begin{figure}
\vskip 0.5in
\epsfxsize = 5.5in
\epsfysize = 5.5in
\centering
\hbox{\raisebox{0.9\epsfysize}{\large $\frac{k'S'(k')}{\xit'}$} \hskip -0.3in
\epsfbox{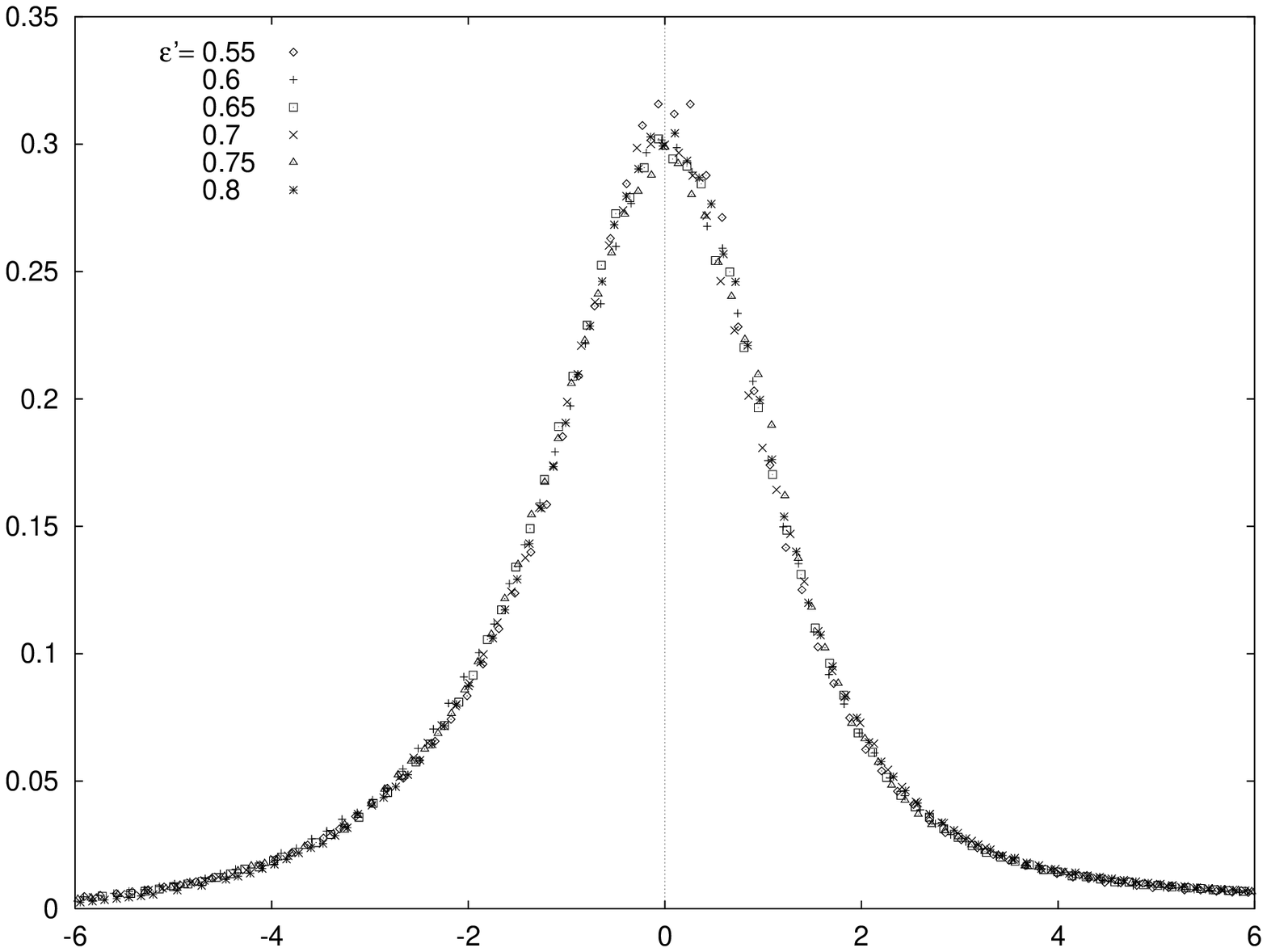}}
$(k'-\km') \xit'$
\caption{}
\lbl{fig_scaling_sdc}
\end{figure}

\newpage
\begin{figure}
\epsfxsize = 6in
\epsfysize = 3in
\centering
\hbox{\raisebox{0.9\epsfysize}{$\xi_2 k^3 S(k)$} \hskip -0.3in
\epsfbox{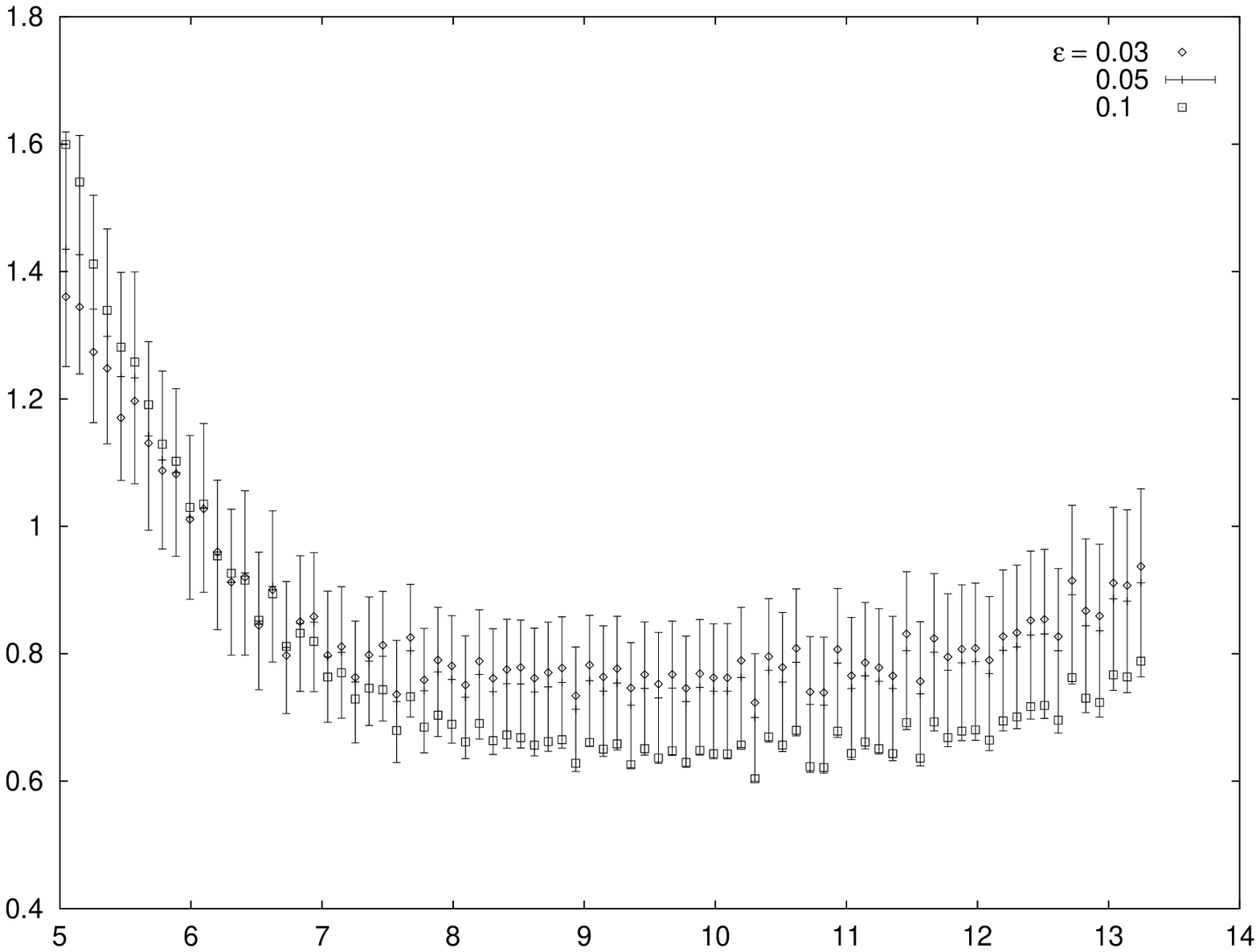}}

\hskip 4in $k$

(a)
\vskip 0.5in
\epsfxsize = 6in
\epsfysize = 3in
\centering
\hbox{\raisebox{0.9\epsfysize}{$\xi'_2 k'^3 S'(k')$} \hskip -0.3in
\epsfbox{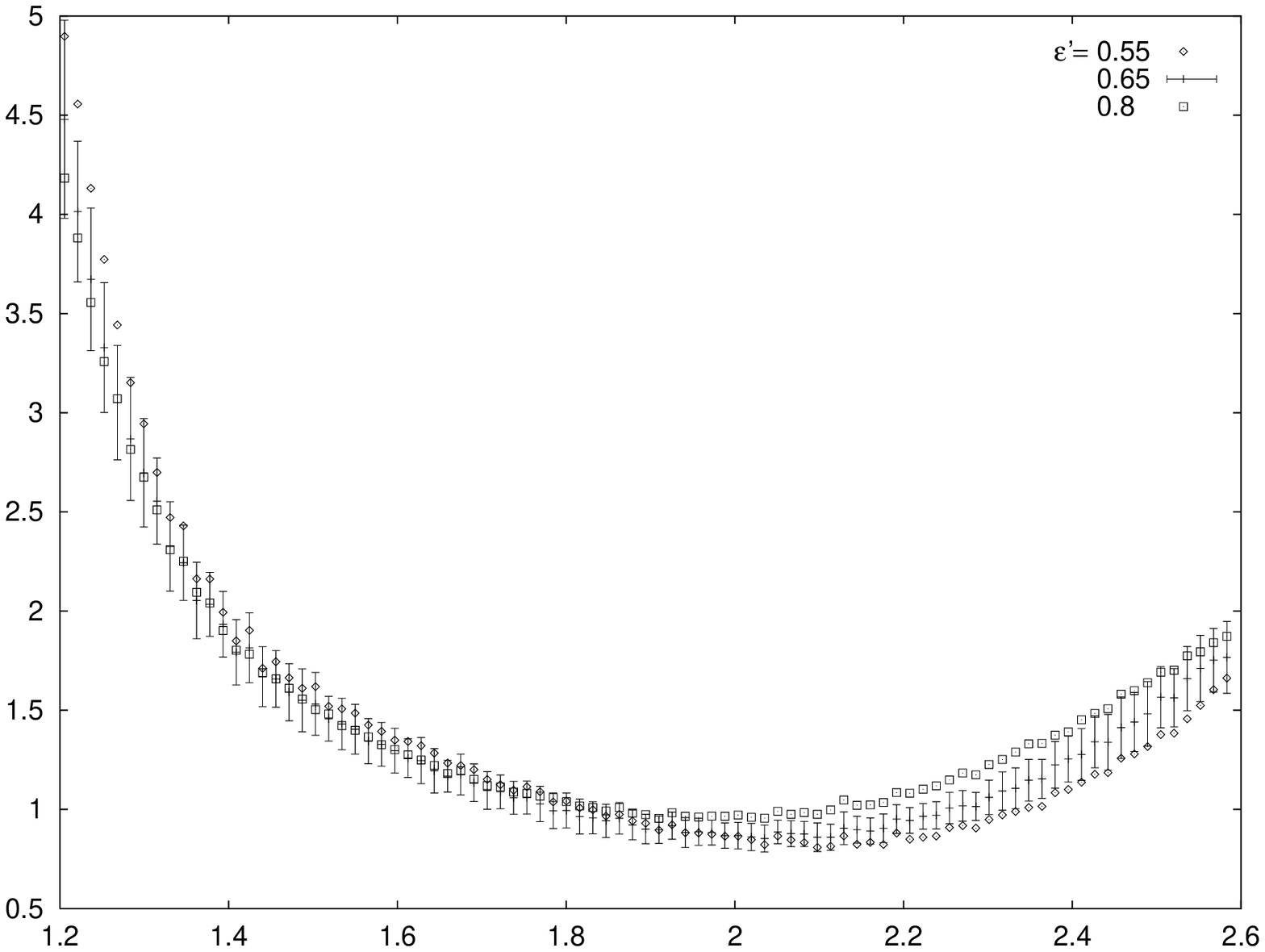}}

\hskip 4in $k'$

(b)
\caption{}
\lbl{fig_porod}
\end{figure}



\end{document}